\documentclass[11pt,a4paper]{article}
\usepackage[]{fontenc}

\usepackage{latexsym}
\usepackage{amsmath,amsthm,amsfonts,amssymb,amscd,ulem}
\usepackage[dvips]{graphicx}

\usepackage{enumerate}

 \theoremstyle{plain}    
 \newtheorem{thm}{Theorem}[section]
 \numberwithin{equation}{section} 
 \numberwithin{figure}{section} 
 \theoremstyle{plain}
 \theoremstyle{definition}
 \newtheorem{defn}[thm]{Definition}
 \theoremstyle{plain}    
 \newtheorem{lem}[thm]{Lemma} 
 \theoremstyle{plain}    
 \newtheorem{cor}[thm]{Corollary} 
 \theoremstyle{remark}
 \newtheorem{rem}[thm]{Remark}
 \theoremstyle{plain}    
 \newtheorem{prop}[thm]{Proposition} 

\makeatletter
 
 \@addtoreset{equation}{section}
\makeatother

\renewcommand{\labelenumi}{(\theenumi)}

\def\rnum#1{\expandafter{\romannumeral #1}}
\def\Rnum#1{\uppercase\expandafter{\romannum}}

\makeatother
\begin{document}
\thispagestyle{empty}

\ 

\vspace{15ex}
\begin{center}\textbf{\huge White noise distribution theory for the
Fermion system}\end{center}{\huge \par}
\vspace{20ex}

\begin{center}{\large Yoshihito Shimada}\end{center}{\large \par}

\begin{center}\textit{Graduate School of Mathematics}\end{center}

\begin{center}\textit{Kyushu University}\end{center}

\begin{center}\textit{1-10-6 Hakozaki, Fukuoka 812-8581}\end{center}

\begin{center}\textit{JAPAN}\end{center}

\vspace{20ex}
\

In this paper, we give the white noise calculus for the Fermion system
and prove the Fock expansion. Each continuous linear operator on Fermionic
white noise functionals is uniquely represented by the series of integral
kernel operators. This series is called the Fock expansion. 

\vspace{1ex}
\hrule
\vspace{1ex}


\texttt{\textbf{\small e-mail: shimada@math.kyushu-u.ac.jp}}{\small \par}

\newpage

\section{Introduction}

In this paper, we give the white noise calculus for the Fermion system
and prove the Fock expansion for any continuous linear operators from
the space of test functionals to the space of generalized functionals.

To the beginning, we mention the motivation of this study. Our white
noise calculus introduced by T. Hida in 1975 is the theory for the
space $\mathcal{E}$ of test functionals and the space $\mathcal{E}^{*}$
of generalized functionals on the infinite dimensional space, for
(continuous) linear operators from $\mathcal{E}$ to $\mathcal{E}^{*}$.
In quantum mechanical physics, the white noise calculus provides us
with a framework of an analysis for the Boson system. (See \cite{Obata-book-1994}.)
In the Boson system, we can represent a continuous linear operator
$\Xi$ from $\mathcal{E}$ to $\mathcal{E}^{*}$ as a series of integral
kernel operators. This representation of $\Xi$ is called the Fock
expansion. The Fock expansion is formulated by T.Hida, N.Obata and
K.Sait{\^o} in \cite{Hida-Obata-Saito}. The Fock expansion is applied
for determining the commutant. For example, N. Obata \cite{Obata-paper-rotation_inv}
used the Fock expansion to obtain the characterization of rotation
invariant operators on white noise functionals. Moreover, in \cite{Shimada},
the author showed irreducibility of the energy representation of a
group of $C^{\infty}$-mappings from a compact Riemann manifold to
a semi-simple compact Lie group. 

As mentioned above, we hope the existence of the Fock expansion for
the Fermion system since the Fock expansion is useful for determining
the commutant. (In fact, we apply the Fock expansion for the Fermion
system to the implementability of Bogoliubov automorphisms of canonical
anti-commutation relations algebra, and we have a partial solution
now. In another paper, we will be able to see an application of the
Fock expansion.) As for the white noise calculus for the Fermion system,
Y. Liao and K. Liu \cite{Liao-Liu} introduced it and showed It{\^o}'s
product formulas for creation, annihilation, and number processes
when the one particle space is $L^{2}(\mathbf{R})$. In this paper,
we define the white noise calculus for the Fermion system when the
one particle space is an abstract Hilbert space. Moreover, we prove
the Fock expansion of continuous linear operators for the Fermion
system. 

Next, we describe an outline of the proof of the Fock expansion for
continuous linear operators on the Fermion Fock space. Since a pair
of Fermions behaves like a Boson, we can show the Fock expansion for
the even part of the Fermion system and extend the result of the even
part to the whole of the Fermion system with the help of the canonical
anti-commutation relations for creation and annihilation operators.

This paper is organized as follows. In section 2, we make the Gelfand
triples for the even part, odd part, and the whole of the Fermion
system and define $S$-transform of generalized white noise functionals
for the even part of the Fermion system. In section 3 and 4 we give
the Fock expansion for the even part of the Fermion system. In section
5, we extend the Fock expansion obtained in section3 to the whole
of the Fermion system.

\section{$S$-transform}

In this section, we make Gelfand triples for the Fermion system and
define the $S$-transform of generalized white noise functionals for
even part of the Fermion system. 

\begin{defn}
\label{thm:property_of_self-adj_op_A}Let $H$ be a complex Hilbert
space with an inner product $(\cdot,\cdot)_{0}$. Let $A$ be a self-adjoint
operator defined on a dense domain $D(A)$. Let $\{\lambda_{j}\}_{j\in\mathbf{N}}$
be eigenvalues of $A$ and $\{ e_{j}\}_{j\in\mathbf{N}}$ be normalized
eigenvectors for $\{\lambda_{j}\}_{j\in\mathbf{N}}$, i.e., $Ae_{j}=\lambda_{j}e_{j}$
, $j\in\mathbf{N}$. Moreover, we also assume the following two conditions
:\renewcommand{\theenumi}{\roman{enumi}}

\renewcommand{\labelenumi}{\rm(\theenumi)}
\begin{enumerate}
\item $\{ e_{j}\}_{j\in\mathbf{N}}$ is a C.O.N.S. of $H$, 
\item Multiplicity of $\{\lambda_{j}\}_{j\in\mathbf{N}}$ is finite and
$1<\lambda_{1}\leq\lambda_{2}\leq\ldots\rightarrow\infty$. 
\end{enumerate}
Then we have the following properties.\renewcommand{\theenumi}{\arabic{enumi}}

\renewcommand{\labelenumi}{\rm(\theenumi)}
\begin{enumerate}
\item For $p\in\mathbf{R}_{\geq0}$ and $x,\, y\in D(A^{p})$, let $(x,y)_{p}:=(A^{p}x,A^{p}y)_{0}$.
Then $(\cdot,\cdot)_{p}$ is an inner product on $D(A^{p})$. Moreover,
$D(A^{p})$ is complete with respect to the norm $|\cdot|_{p}$, that
is, the pair $E_{p}:=(D(A^{p}),|\cdot|_{p})$ is a Hilbert space. 
\item For $q\geq p\geq0$, let $j_{p,q}:E_{q}\hookrightarrow E_{p}$ be
the inclusion map. Then every inclusion map is continuous and has
a dense image. Then $\{ E_{p},j_{p,q}\}$ is a reduced projective
system. 
\item A standard countable Hilbert space\[
E:=\lim_{\leftarrow}E_{p}=\bigcap_{p\geq0}E_{p}\]
constructed from the pair $(H,A)$ is a reflexive Fr\'echet space.
We call $E$ a CH-space simply.
\item From (3), we have ${\displaystyle E^{*}=\lim_{\rightarrow}E_{p}^{*}}$
as a topological vector space, i.e. the strong topology on $E^{*}$
and the inductive topology on ${\displaystyle \lim_{\rightarrow}E_{p}^{*}}$
coincide.
\item Let $p\in\mathbf{R}_{\geq0}$ and $(x,y)_{-p}:=(A^{-p}x,A^{-p}y)_{0}$.
Then $(\cdot,\cdot)_{-p}$ is an inner product on $H$. 
\item Let $E_{-p}$ be the completion of $H$ with respect to the norm $|\cdot|_{-p}$.
For $q\geq p\geq0$, we can consider the inclusion map $i_{-q,-p}:E_{-p}\hookrightarrow E_{-q}$.
Then $\{ E_{-p},i_{-q,-p}\}$ is an inductive system. Moreover, $E_{-p}$
and $E_{p}^{*}$ are anti-linear isomorphic and isometric. Thus, from
(4), we have\[
E^{*}=\lim_{\rightarrow}E_{-p}=\bigcup_{p\geq0}E_{-p}.\]
 
\end{enumerate}
\end{defn}
Furthermore, we require for the operator $A$ that there exists $\alpha>0$
such that $A^{-\alpha}$ is a Hilbert-Schmidt class operator, namely\begin{equation}
\delta^{2}:=\sum_{j=1}^{\infty}\lambda_{j}^{-2\alpha}<\infty.\label{eq:Hilbert-Schmidt_inverse}\end{equation}
 From this condition, $E$ (resp. $E^{*}$) is a nuclear space. Thus
we can define the $\pi$-tensor topology $E\otimes_{\pi}E$ (resp.
$E^{*}\otimes_{\pi}E^{*}$) of $E$ (resp. $E^{*}$). If there is
no danger of confusion, we will use the notation $E\otimes E$ (resp.
$E^{*}\otimes E^{*}$) simply.

We denote the canonical bilinear form on $E^{*}\times E$ by $\left\langle \cdot,\cdot\right\rangle $.
We have the following natural relation between the canonical bilinear
form on $E^{*}\times E$ and the inner product on $H$ :\[
\left\langle f,g\right\rangle =(Jf,g)_{0}\]
for all $f\in H$ and $g\in E$. $Jf\in H$ is the complex conjugate
of $f\in H$.

\begin{defn}
\label{thm:def_of_alternizer}Let $X$ be a Hilbert space, or a CH-space.
\begin{enumerate}
\item Let $g_{1}$, $\ldots$ , $g_{n}\in X$. We define the anti-symmetrization
$\mathcal{A}_{n}(g_{1}\otimes\ldots\otimes g_{n})$ of $g_{1}\otimes\ldots\otimes g_{n}\in X^{\otimes n}$
as follows.\[
\mathcal{A}_{n}(g_{1}\otimes\ldots\otimes g_{n}):=g_{1}\wedge\ldots\wedge g_{n}:=\frac{1}{n!}\sum_{\sigma\in\mathfrak{S}_{n}}\mathrm{sign}(\sigma)g_{\sigma(1)}\otimes\ldots\otimes g_{\sigma(n)},\]
where $\mathfrak{S}_{n}$ is the set of all permutations of $\{1,2,\ldots,n\}$. 
\item If $f\in X^{\otimes n}$ satisfies $\mathcal{A}_{n}(f)=f$, then we
call $f$ anti-symmetric. We denote the set of all anti-symmetric
elements of $X^{\otimes n}$ by $X^{\wedge n}$ and we call $X^{\wedge n}$
the $n$-th anti-symmetric tensor of $X$. If $X$ is a Hilbert space,
then $\mathcal{A}_{n}$ is a projection from $X^{\otimes n}$ to $X^{\wedge n}$
.
\item Let $X$ be a CH-space. For $F\in(X^{\otimes n})^{*}$ and $\sigma\in\mathfrak{S}_{n}$,
let $F^{\sigma}$ be an element of $(X^{\otimes n})^{*}$ satisfying\[
\left\langle F^{\sigma},g_{1}\otimes\ldots\otimes g_{n}\right\rangle :=\left\langle F,g_{\sigma^{-1}(1)}\otimes\ldots\otimes g_{\sigma^{-1}(n)}\right\rangle ,\quad g_{i}\in X.\]
Then we define the anti-symmetrization $\mathcal{A}_{n}(F)$ as follows.\[
\mathcal{A}_{n}(F):=\frac{1}{n!}\sum_{\sigma\in\mathfrak{S}_{n}}\mathrm{sign}(\sigma)F^{\sigma}.\]

\item If $F\in(X^{\otimes n})^{*}$ satisfies $\mathcal{A}_{n}(F)=F$, we
call $F$ anti-symmetric. We denote the set of all anti-symmetric
elements of $(X^{\otimes n})^{*}$ by $(X^{\wedge n})^{*}$.
\end{enumerate}
\end{defn}
From the above discussion, we obtain a Gelfand triple :\[
E\subset H\subset E^{*}.\]

\begin{lem}
Let $H$ be a Hilbert space. 
\begin{enumerate}
\item Let \[
(f_{1}\otimes\ldots\otimes f_{n},g_{1}\otimes\ldots\otimes g_{n})_{0}:=(f_{1},g_{1})_{0}\ldots(f_{n},g_{n})_{0}\]
for $f_{i}$, $g_{j}\in H$, $i,j=1,2,\ldots,n$. Then\[
(f_{1}\wedge\ldots\wedge f_{n},g_{1}\wedge\ldots\wedge g_{n})_{0}=\frac{1}{n!}\det\left((f_{i},g_{j})_{0}\right)_{1\leq i,j\leq n}.\]
Moreover $\mathcal{A}_{n}$ is a projection with respect to $(\cdot,\cdot)_{0}$.
\item For $f\in H^{\wedge n}$ and $g\in H^{\wedge m}$, we have\[
\left|f\wedge g\right|_{0}\leq\left|f\right|_{0}\left|g\right|_{0}.\]

\end{enumerate}
\end{lem}
Next, we define the Fermion Fock space and the second quantization
of a linear operator.

\begin{defn}
Let $H$ be a Hilbert space and $A$ be a linear operator on $H$.
\begin{enumerate}
\item Let\begin{gather*}
\Gamma(H):=\left\{ \sum_{n=0}^{\infty}\phi_{n}\,|\,\phi_{n}\in H^{\wedge n},\,\left\Vert \sum_{n=0}^{\infty}\phi_{n}\right\Vert _{0}^{2}:=\sum_{n=0}^{\infty}n!\left|\phi_{n}\right|_{0}^{2}<+\infty\right\} ,\\
\left(\sum_{n=0}^{\infty}\phi_{n},\sum_{n=0}^{\infty}\psi_{n}\right)_{0}=\sum_{n\in\mathbf{Z}_{\geq0}}n!(\phi_{n},\psi_{n})_{0}.\end{gather*}
Then we call $\Gamma(H)$ the Fermion Fock space. The Fermion Fock
space $\Gamma(H)$ is a Hilbert space with respect to the inner product
$(\cdot,\cdot)_{0}$. Moreover let\begin{gather*}
\Gamma^{+}(H):=\left\{ \sum_{n=0}^{\infty}\phi_{2n}\,|\,\phi_{2n}\in H^{\wedge(2n)},\,\sum_{n=0}^{\infty}(2n)!\left|\phi_{2n}\right|_{0}^{2}<+\infty\right\} ,\\
\Gamma^{-}(H):=\left\{ \sum_{n=0}^{\infty}\phi_{2n+1}\,|\,\phi_{2n+1}\in H^{\wedge(2n+1)},\,\sum_{n=0}^{\infty}(2n+1)!\left|\phi_{2n+1}\right|_{0}^{2}<+\infty\right\} .\end{gather*}
Then we call $\Gamma^{+}(H)$ (resp. $\Gamma^{-}(H)$) the even part
of the Fermion Fock space (resp. the odd part of the Fermion Fock
space).
\item We call \[
\Gamma(A):=\sum_{n=0}^{\infty}A^{\otimes n}\]
the second quantization of $A$. Let\[
\Gamma^{+}(A):=\Gamma(A)|\Gamma^{+}(H),\quad\Gamma^{-}(A):=\Gamma(A)|\Gamma^{-}(H).\]

\end{enumerate}
\end{defn}
~

\begin{defn}
\label{thm:def_of_CH-space_Gelfand_triple}Let $H$ be a complex Hilbert
space and $A$ be a self-adjoint operator on $H$ satisfying the conditions
(i) and (ii) in Lemma \ref{thm:property_of_self-adj_op_A} and \eqref{eq:Hilbert-Schmidt_inverse}.
Then we can define a CH-space $\mathcal{E}$ constructed from $(\Gamma(H),\Gamma(A))$
and we obtain a Gelfand triple :\[
\mathcal{E}\subset\Gamma(H)\subset\mathcal{E}^{*}.\]
Moreover, let $\mathcal{E}_{+}$(resp. $\mathcal{E}_{-}$) be a CH-space
constructed from $(\Gamma^{+}(H),\Gamma^{+}(A))$ (resp. $(\Gamma^{-}(H),\Gamma^{-}(A))$)
and we obtain Gelfand triples :\[
\mathcal{E}_{+}\subset\Gamma^{+}(H)\subset\mathcal{E}_{+}^{*},\quad\mathcal{E}_{-}\subset\Gamma^{-}(H)\subset\mathcal{E}_{-}^{*}.\]
Then an element of $\mathcal{E}$ (or $\mathcal{E}_{+}$, $\mathcal{E}_{-}$)
is called a test (white noise) functional and an element of $\mathcal{E}^{*}$(or
$\mathcal{E}_{+}^{*}$, $\mathcal{E}_{-}^{*}$) is called a generalized
(white noise) functional.
\end{defn}
\begin{cor}
Let $\phi:=\sum_{n=0}^{\infty}\phi_{n}\in\Gamma(H)$, $\phi_{n}\in H^{\wedge n}$.
Then $\phi\in\mathcal{E}$ if and only if $\phi_{n}\in E^{\wedge n}$
for all $n\geq0$. Moreover, it holds that\[
\left\Vert \phi\right\Vert _{p}^{2}:=\left\Vert \Gamma(A)^{p}\phi\right\Vert _{0}^{2}=\sum_{n=0}^{\infty}n!|\phi_{n}|_{p}^{2}<+\infty\]
for all $p\geq0$. We can also show this statement in case of $\phi\in\Gamma^{+}(H)$
and $\phi\in\Gamma^{-}(H)$. 
\end{cor}
\begin{rem}
Let $H$ be a Hilbert space. Then $\zeta\wedge\eta=\eta\wedge\zeta$
for $\zeta$, $\eta\in H^{\wedge2}$. Thus we can define $\zeta^{\wedge n}$,
$n\geq0$ for all $\zeta\in H^{\wedge2}$. This shows that a pair
of Fermions behaves like a Boson. 
\end{rem}
\begin{defn}
For any $\zeta\in H^{\wedge2}$, we define an element $e^{+}(\zeta)\in\Gamma^{+}(H)$
as follows :\[
e^{+}(\zeta):=\sum_{n=0}^{\infty}\frac{1}{(2n)!}\zeta^{\wedge n}.\]
 We can check the well-definedness of $e^{+}(\zeta)$ easily.
\end{defn}
\begin{cor}
If $\zeta\in E^{\wedge2}$, then $e^{+}(\zeta)\in\mathcal{E}_{+}$.
\end{cor}
\begin{proof}
Since\[
|\phi\otimes\psi|_{p}\leq|\phi|_{p}|\psi|_{p}\]
for all $\phi\in E^{\otimes l}$, $\psi\in E^{\otimes m}$, and $p\geq0$
(see Lemma \ref{thm:estimations_for_contraction_(otimes)} ), we have\[
|\zeta^{\wedge n}|_{p}\leq|\zeta|_{p}^{n}\]
for all $\zeta\in E^{\wedge2}$, $p\geq0$. Thus\[
\left\Vert e^{+}(\zeta)\right\Vert _{p}^{2}=\sum_{n=0}^{\infty}(2n)!\left|\frac{1}{(2n)!}\zeta^{\wedge n}\right|_{p}^{2}\leq\sum_{n=0}^{\infty}\frac{1}{(2n)!}|\zeta|_{p}^{2n}\leq\sum_{n=0}^{\infty}\frac{1}{n!}|\zeta|_{p}^{2n}=\exp(|\zeta|_{p}^{2}).\]
This implies $e^{+}(\zeta)\in\mathcal{E}_{+}$.
\end{proof}
\begin{prop}
Let $H$ be a Hilbert space and $S:=\{ e^{+}(\zeta)|\zeta\in H^{\wedge2}\}$.
Then $\mathrm{span}_{\mathbf{C}}[S]$ is a dense subspace of $\Gamma^{+}(H)$
with respect to the norm $\left\Vert \cdot\right\Vert _{0}$. Moreover,
for $S':=\{ e^{+}(\zeta)|\zeta\in E^{\wedge2}\}$, $\mathrm{span}_{\mathbf{C}}[S']$
is a dense subspace of $\mathcal{E}_{+}$ with respect to the topology
of $\mathcal{E}_{+}$.
\end{prop}
\begin{proof}
We prove that $\mathrm{span}_{\mathbf{C}}[S']$ is a dense subspace
of $\mathcal{E}_{+}$ first. Let $\{ e_{i}\}_{i=1}^{\infty}$ be a
C.O.N.S. of $H$ satisfying the conditions (i) and (ii) in Lemma \ref{thm:property_of_self-adj_op_A}
and \eqref{eq:Hilbert-Schmidt_inverse}. For \[
\zeta:=e_{i_{1}}\wedge e_{i_{2}}+\ldots+e_{i_{2n-1}}\wedge e_{i_{2n}}\in E^{\wedge2},\]
we have\[
\zeta^{\wedge n}=n!\, e_{i_{1}}\wedge e_{i_{2}}\wedge\ldots\wedge e_{i_{2n}}.\]
We note that $\{(\lambda_{i_{1}}\ldots\lambda_{i_{2n}})^{-p}e_{i_{1}}\wedge\ldots\wedge e_{i_{2n}}\,|\, i_{1}<\ldots<i_{2n}\}$
is a C.O.N.S. of a Hilbert space $E_{p}^{\wedge n}$. This implies
that $\mathrm{span}_{\mathbf{C}}[\{\zeta^{\wedge n}|\zeta\in E^{\wedge2}\}]$
is a dense subspace of $E_{p}^{\wedge(2n)}$ with respect to the norm
$|\cdot|_{p}$ for any $p\geq0$, i.e., \begin{equation}
E_{p}^{\wedge(2n)}=(E_{p}^{\wedge2})^{\wedge n}=\overline{\mathrm{span}}_{\mathbf{C}}[\{\zeta^{\wedge n}|\zeta\in E^{\wedge2}\}].\label{eq:H^(2n)_generated_by_zeta^n}\end{equation}
On the other hand, we can check that $\zeta^{\wedge n}$, $\zeta\in E^{\wedge2}$
are elements of the closure of $\mathrm{span}_{\mathbf{C}}[S']$ with
respect to the norm $|\cdot|_{p}$ for any $p\geq0$, i.e.,\begin{equation}
\zeta^{\wedge n}\in\overline{\mathrm{span}}_{\mathbf{C}}[S'].\label{eq:zeta_in_span(S)}\end{equation}
We show \eqref{eq:zeta_in_span(S)} by induction on $n\geq0$. For $n=0$,
\eqref{eq:zeta_in_span(S)} follows from\[
\zeta^{\wedge0}=1=e^{+}(0)\in S'.\]
 Now we assume \eqref{eq:zeta_in_span(S)} for $n=1,2,\ldots,r$. Then
$\zeta^{\wedge(r+1)}\in\overline{\mathrm{span}}_{\mathbf{C}}[S']$
follows from \[
\zeta^{\wedge(r+1)}=(2r+2)!\,\left\Vert \cdot\right\Vert _{p}\frac{\,\,}{\,\,}\lim_{\varepsilon\rightarrow0}\frac{1}{\varepsilon^{r+1}}\left(e^{+}(\varepsilon\zeta)-\sum_{n=0}^{r}\frac{1}{(2n)!}(\varepsilon\zeta)^{\wedge n}\right).\]
Thus we have \eqref{eq:zeta_in_span(S)} for all $n\geq0$. \eqref{eq:H^(2n)_generated_by_zeta^n}
and \eqref{eq:zeta_in_span(S)} imply\[
\overline{\mathrm{span}}_{\mathbf{C}}[S']=\overline{\mathrm{span}}_{\mathbf{C}}[\{\zeta^{\wedge n}|\zeta\in E_{+},n\in\mathbf{Z}_{\geq0}\}]=(D(\Gamma^{+}(A)^{p}),\,\left\Vert \cdot\right\Vert _{p})\]
for all $p\geq0$. Thus $\mathrm{span}_{\mathbf{C}}[S']$ is a dense
subspace of $\mathcal{E}_{+}$ with respect to the topology of $\mathcal{E}_{+}$.

In the same manner, we can show that $\mathrm{span}_{\mathbf{C}}[S]$
is a dense subspace of $\Gamma^{+}(H)$ with respect to the norm $\left\Vert \cdot\right\Vert _{0}$. 
\end{proof}
\begin{defn}
For $\Phi\in\mathcal{E}_{+}^{*}$, we define a function $S\Phi$ on
$E^{\wedge2}$ as follows :\[
(S\Phi)(\zeta):=\left\langle \!\left\langle \Phi,e^{+}(\zeta)\right\rangle \!\right\rangle ,\quad\zeta\in E^{\wedge2}.\]
Then we call $S\Phi$ the $S$-\textit{transform} of $\Phi$.
\end{defn}
\begin{cor}
Let $\Phi=(\Phi_{2n})_{n=0}^{\infty}\in\mathcal{E}_{+}^{*}$. Then\[
(S\Phi)(\zeta)=\sum_{n=0}^{\infty}\left\langle \Phi_{2n},\zeta^{\wedge n}\right\rangle ,\quad\zeta\in E^{\wedge2}\]
and the right hand side converges absolutely.
\end{cor}
\begin{proof}
If $p\geq0$ satisfies $\left\Vert \Phi\right\Vert _{-p}<+\infty$,
then\begin{align*}
\sum_{n=0}^{\infty}\left|\left\langle \Phi_{2n},\zeta^{\wedge n}\right\rangle \right| & \leq\sum_{n=0}^{\infty}\sqrt{(2n)!}\left|\Phi_{2n}\right|_{-p}\frac{1}{\sqrt{(2n)!}}|\zeta^{\wedge n}|_{p}\\
 & \leq\left(\sum_{n=0}^{\infty}(2n)!\left|\Phi_{2n}\right|_{-p}^{2}\right)^{\frac{1}{2}}\left(\sum_{n=0}^{\infty}\frac{1}{(2n)!}|\zeta|_{p}^{2n}\right)^{\frac{1}{2}}\\
 & \leq\left\Vert \Phi\right\Vert _{-p}\exp\left(\frac{1}{2}|\zeta|_{p}^{2}\right)<+\infty.\end{align*}

\end{proof}
\begin{prop}
\label{thm:S-trans_is_holomorphic}For any $\zeta,\eta\in E^{\wedge2}$
and $\Phi\in\mathcal{E}_{+}^{*}$, a function\[
\mathbf{C}\ni z\mapsto S\Phi(z\zeta+\eta)\in\mathbf{C}\]
is holomorphic in $\mathbf{C}$.
\end{prop}
\begin{proof}
Let $\Phi=(\Phi_{2n})_{n=0}^{\infty}$. Then\begin{align*}
(S\Phi)(z\zeta+\eta) & =\sum_{n=0}^{\infty}\left\langle \Phi_{2n},(z\zeta+\eta)^{\wedge n}\right\rangle \\
 & =\sum_{n=0}^{\infty}\sum_{k=0}^{n}\left(\begin{array}{c}
n\\
k\end{array}\right)z^{k}\left\langle \Phi_{2n},\zeta^{\wedge k}\wedge\eta^{\wedge(n-k)}\right\rangle \\
 & =\sum_{k=0}^{\infty}\left(\sum_{n=0}^{\infty}\left(\begin{array}{c}
n+k\\
k\end{array}\right)\left\langle \Phi_{2(n+k)},\zeta^{\wedge k}\wedge\eta^{\wedge n}\right\rangle \right)z^{k}.\end{align*}
Now let\[
a_{k}:=\sum_{n=0}^{\infty}\left(\begin{array}{c}
n+k\\
k\end{array}\right)\left\langle \Phi_{2(n+k)},\zeta^{\wedge k}\wedge\eta^{\wedge n}\right\rangle \]
and we show that the radius of convergence $R$ is infinite, that
is, \[
\frac{1}{R}=\limsup_{k\mapsto\infty}|a_{k}|^{\frac{1}{k}}=0.\]
 $|a_{k}|$ satisfies\begin{align*}
|a_{k}| & \leq\sum_{n=0}^{\infty}\frac{(n+k)!}{n!k!}|\Phi_{2(n+k)}|_{-p}|\zeta|_{p}^{k}|\eta|_{p}^{n}\\
 & =\frac{|\zeta|_{p}^{k}}{k!}\sum_{n=0}^{\infty}\sqrt{(n+k)!}\,|\Phi_{2(n+k)}|_{-p}\frac{\sqrt{(n+k)!}}{n!}|\eta|_{p}^{n}\\
 & \leq\frac{|\zeta|_{p}^{k}}{k!}\left(\sum_{n=0}^{\infty}(n+k)!|\Phi_{2(n+k)}|_{-p}^{2}\right)^{\frac{1}{2}}\left(\sum_{n=0}^{\infty}\frac{(n+k)!}{n!n!}|\eta|_{p}^{2n}\right)^{\frac{1}{2}}\\
 & \leq\frac{|\zeta|_{p}^{k}}{k!}\left(\sum_{n=0}^{\infty}(2n+2k)!|\Phi_{2(n+k)}|_{-p}^{2}\right)^{\frac{1}{2}}\left(\sum_{n=0}^{\infty}\frac{(n+k)!}{n!n!}|\eta|_{p}^{2n}\right)^{\frac{1}{2}}\\
 & =\frac{|\zeta|_{p}^{k}}{k!}\left\Vert \Phi\right\Vert _{-p}\left(\sum_{n=0}^{\infty}\frac{(n+k)!}{n!n!}|\eta|_{p}^{2n}\right)^{\frac{1}{2}}.\end{align*}
Since we have\[
\sum_{n=0}^{\infty}\frac{(n+k)!}{n!n!}t^{n}\leq(t+k)^{k}\exp t\]
for all $t\geq0$ and $k\in\mathbf{Z}_{\geq0}$(see Lemma 3.2.9 of
\cite{Obata-book-1994}), it holds that\[
|a_{k}|\leq\frac{|\zeta|_{p}^{k}}{k!}\left\Vert \Phi\right\Vert _{-p}(|\eta|_{p}^{2}+k)^{\frac{k}{2}}\exp\left(\frac{1}{2}|\eta|_{p}^{2}\right).\]
If we note that $k^{k}/k!\leq\exp k$, i.e., \[
\frac{1}{k!}\leq\left(\frac{1}{k}\right)^{k}\exp(k),\]
it holds that\[
|a_{k}|^{\frac{1}{k}}\leq|\zeta|_{p}\left\Vert \Phi\right\Vert _{-p}^{\frac{1}{k}}\frac{\sqrt{|\eta|_{p}^{2}+k}}{k}\exp\left(1+\frac{1}{2k}|\eta|_{p}\right)\rightarrow0\,(k\rightarrow\infty).\]

\end{proof}

\section{\label{sec:Fock expansion (even)}Fock expansion for the even part
of the fermion system}

In order to discuss integral kernel operators, we define a contraction
of tensor product.

\begin{defn}
Let $H$ be a complex Hilbert space and $A$ be a self-adjoint operator
on $H$ satisfying the conditions (i) and (ii) in Lemma \ref{thm:property_of_self-adj_op_A}
and \eqref{eq:Hilbert-Schmidt_inverse}. Let\[
e(\mathbf{i}):=e_{i_{1}}\otimes\ldots\otimes e_{i_{l}},\quad\mathbf{i}:=(i_{1},\ldots,i_{l})\in\mathbf{N}^{l}.\]

\begin{enumerate}
\item Let $l$, $m\in\mathbf{N}$. For $F\in\left(E^{\otimes(l+m)}\right)^{*}$,
let\[
|F|_{l,m;p,q}^{2}:=\sum_{\mathbf{i},\mathbf{j}}\left|\left\langle F,e(\mathbf{i})\otimes e(\mathbf{j})\right\rangle \right|^{2}\left|e(\mathbf{i})\right|_{p}^{2}\left|e(\mathbf{j})\right|_{q}^{2}\]
where $\mathbf{i}$ and $\mathbf{j}$ run the whole $\mathbf{N}^{l}$
and $\mathbf{N}^{m}$ respectively.
\item Let $l$, $n\in\mathbf{N}$ and $m\in\mathbf{Z}_{\geq0}$. For $F\in\left(E^{\otimes(l+m)}\right)^{*}$
and $g\in E^{\otimes(m+n)}$, we define a left contraction $F\otimes^{m}g\in\left(E^{\otimes(l+n)}\right)^{*}$
of $F$ and $g$ as follows.\[
F\otimes^{m}g:=\sum_{\mathbf{j},\mathbf{k}}\left(\sum_{\mathbf{i}}\left\langle F,e(\mathbf{i})\otimes e(\mathbf{j})\right\rangle \left\langle g,e(\mathbf{i})\otimes e(\mathbf{k})\right\rangle \right)e(\mathbf{j})\otimes e(\mathbf{k})\]
where $\mathbf{i}$, $\mathbf{j}$, and $\mathbf{k}$ run the whole
$\mathbf{N}^{m}$, $\mathbf{N}^{l}$, and $\mathbf{N}^{n}$ respectively.
Similarly, we define a right contraction $F\otimes_{m}g\in\left(E^{\otimes(l+n)}\right)^{*}$
of $F$ and $g$ as follows.\[
F\otimes_{m}g:=\sum_{\mathbf{j},\mathbf{k}}\left(\sum_{\mathbf{i}}\left\langle F,e(\mathbf{j})\otimes e(\mathbf{i})\right\rangle \left\langle g,e(\mathbf{k})\otimes e(\mathbf{i})\right\rangle \right)e(\mathbf{j})\otimes e(\mathbf{k})\]
where $\mathbf{i}$, $\mathbf{j}$, and $\mathbf{k}$ run the whole
$\mathbf{N}^{m}$, $\mathbf{N}^{l}$, and $\mathbf{N}^{n}$ respectively. 
\end{enumerate}
\end{defn}
We check the well-definedness of the contraction. Let $\rho$ be the
operator norm of $A^{-1}$, that is, $\rho$ is the inverse of the
first eigenvalue of $A$. We remark that \[
|e(\mathbf{i})|_{-p}|e(\mathbf{i})|_{p}=1\]
and\[
|e(\mathbf{i})|_{p}\leq\rho^{nr}|e(\mathbf{i})|_{p+r}\]
for all $p\in\mathbf{R}$, $r\geq0$ and $\mathbf{i}\in\mathbf{N}^{n}$.
Then we have\begin{align*}
|F & \otimes^{m}g|_{p}^{2}\\
 & =|F\otimes^{m}g|_{l,n;p,p}^{2}\\
 & =\sum_{\mathbf{j},\mathbf{k}}\left|\sum_{\mathbf{i}}\left\langle F,e(\mathbf{i})\otimes e(\mathbf{j})\right\rangle \left\langle g,e(\mathbf{i})\otimes e(\mathbf{k})\right\rangle |e(\mathbf{i})|_{-q}|e(\mathbf{i})|_{q}\right|^{2}\left|e(\mathbf{j})\right|_{p}^{2}\left|e(\mathbf{k})\right|_{p}^{2}\\
 & \leq\sum_{\mathbf{j},\mathbf{k}}\left(\sum_{\mathbf{i}}\left|\left\langle F,e(\mathbf{i})\otimes e(\mathbf{j})\right\rangle \right|^{2}|e(\mathbf{i})|_{-q}^{2}\right)\\
 & \qquad\qquad\times\left(\sum_{\mathbf{i}}\left|\left\langle g,e(\mathbf{i})\otimes e(\mathbf{k})\right\rangle \right|^{2}\rho^{2mr}|e(\mathbf{i})|_{q+r}^{2}\right)\left|e(\mathbf{j})\right|_{p}^{2}\rho^{2nr}\left|e(\mathbf{k})\right|_{p+r}^{2}\\
 & =\rho^{2(m+n)r}|F|_{m,l;-q,p}^{2}|g|_{m,n;q+r,p+r}^{2}\\
 & \leq\rho^{2(m+n)r}|F|_{m,l;-q,p}^{2}|g|_{\mathrm{max}\{ p,q\}+r}^{2}.\end{align*}
for $F\in(E^{\otimes(l+m)})^{*}$ and $g\in E^{\otimes(m+n)}$.Therefore
$F\otimes_{m}g\in\left(E^{\otimes(l+n)}\right)^{*}$ and we obtain

\begin{lem}
\label{thm:estimations_for_contraction_(otimes)} Let $F\in(E^{\otimes(l+m)})^{*}$
and $g\in E^{\otimes(m+n)}$. Then\begin{gather*}
|F\otimes^{m}g|_{p}\leq\rho^{(m+n)r}|F|_{m,l;-q,p}|g|_{\mathrm{max}\{ p,q\}+r},\\
|F\otimes_{m}g|_{p}\leq\rho^{(m+n)r}|F|_{l,m;p,-q}|g|_{\mathrm{max}\{ p,q\}+r}\end{gather*}
 for any $p\in\mathbf{R}$, $q\in\mathbf{R}$, and $r\geq0$. 
\end{lem}
The following lemma is easily checked.

\begin{lem}
For $F\in(E^{\otimes(l+m)})^{*}$ and $g\in E^{\otimes(l+n)}$, put\begin{gather*}
F\wedge^{m}g:=\mathcal{A}_{l+n}(F\otimes^{m}g),\\
F\wedge_{m}g:=\mathcal{A}_{l+n}(F\otimes_{m}g).\end{gather*}
 Then $F\wedge^{m}g$ and $F\wedge_{m}g$ are elements of $(E^{\wedge(l+n)})^{*}$.
If $F\in(E^{\wedge(l+m)})^{*}$ and $g\in E^{\wedge(l+n)},$then\[
F\wedge^{m}g=(-1)^{m(l+n)}F\wedge_{m}g.\]
 Thus the left contraction $F\wedge^{m}g$ coincides with the right
contraction $F\wedge_{m}g$ if $m$ is an even number.
\end{lem}
Since a map $\mathcal{A}_{n}:E^{\otimes n}\rightarrow E^{\wedge n}$
is a projection commuting with $A^{\otimes n}$, we can show the following
lemma easily.

\begin{lem}
\label{lem:estimation for contraction (wedge)}Let $F\in(E^{\wedge(l+m)})^{*}$,
$g\in E^{\wedge(m+n)}$. Then\begin{gather*}
|F\wedge^{m}g|_{p}\leq\rho^{(m+n)r}|F|_{m,l;-q,p}|g|_{\mathrm{max}\{ p,q\}+r},\\
|F\wedge_{m}g|_{p}\leq\rho^{(m+n)r}|F|_{l,m;p,-q}|g|_{\mathrm{max}\{ p,q\}+r}\end{gather*}
for any $p\in\mathbf{R}$, $q\in\mathbf{R}$, and $r\geq0$. 
\end{lem}
Put\[
t_{l,m}(\psi)=\sum_{\mathbf{i},\mathbf{j}}\langle\psi,e(\mathbf{i})\otimes e(\mathbf{j})\rangle e(\mathbf{j})\otimes e(\mathbf{i}),\]
where $\mathbf{i}$, $\mathbf{j}$ run the whole $\mathbf{N}^{l}$
and $\mathbf{N}^{m}$ respectively. 

\begin{lem}
Let $\kappa\in(E^{\otimes(l+m)})^{*}$, $\psi\in E^{\otimes(l+n)}$
and $\phi\in E^{\otimes(m+n)}$. Then\[
\langle\kappa\otimes_{m}\phi,\psi\rangle=\langle\kappa,t_{l,n}(\psi)\otimes^{n}\phi\rangle.\]

\end{lem}
\begin{proof}
It is easily checked that\begin{align*}
\langle(e(\mathbf{i}) & \otimes e(\mathbf{i}'))\otimes_{m}(e(\mathbf{j})\otimes e(\mathbf{j}')),\, e(\mathbf{k})\otimes e(\mathbf{k}')\rangle\\
 & =\langle e(\mathbf{i})\otimes e(\mathbf{i}'),\,(e(\mathbf{k}')\otimes e(\mathbf{k}))\otimes^{n}(e(\mathbf{j})\otimes e(\mathbf{j}'))\rangle\end{align*}
for $\mathbf{i},\,\mathbf{k}\in\mathbf{N}^{l}$, $\mathbf{i}',\,\mathbf{j}'\in\mathbf{N}^{m}$,
$\mathbf{j},\,\mathbf{k}'\in\mathbf{N}^{n}$. Thus\begin{align*}
\langle\kappa & \otimes_{m}\phi,\psi\rangle\\
 & :=\sum\langle\kappa,e(\mathbf{i})\otimes e(\mathbf{i}')\rangle\langle\phi,e(\mathbf{j})\otimes e(\mathbf{j}')\rangle\langle\psi,e(\mathbf{k})\otimes e(\mathbf{k}')\rangle\\
 & \qquad\times\langle(e(\mathbf{i})\otimes e(\mathbf{i}'))\otimes_{m}(e(\mathbf{j})\otimes e(\mathbf{j}')),\, e(\mathbf{k})\otimes e(\mathbf{k}')\rangle\\
 & =\langle\kappa,t_{l,n}(\psi)\otimes^{n}\phi\rangle\end{align*}

\end{proof}
\begin{cor}
For $\kappa\in(E^{\otimes2(l+m)})^{*}$, $\psi\in E^{\wedge2(l+n)}$
and $\phi\in E^{\otimes2(m+n)}$, we have\[
\langle\kappa\otimes_{2m}\phi,\psi\rangle=\langle\kappa,\psi\otimes^{2n}\phi\rangle.\]
Moreover, if $\phi\in E^{\wedge2(m+n)}$, then\[
\langle\kappa\wedge_{2m}\phi,\psi\rangle=\langle\kappa,\psi\otimes^{2n}\phi\rangle.\]

\end{cor}
\begin{proof}
This corollary is easily checked.
\end{proof}
We mention continuity of linear operators on locally convex spaces
before discussing integral kernel operators.

\begin{lem}
Let $X$ and $Y$ be locally convex spaces with seminorms $\{|\cdot|_{X,q}\}_{q\in Q}$
and $\{|\cdot|_{Y,p}\}_{p\in P}$ respectively. Let $\mathcal{L}(X,Y)$
be the set of all continuous linear operators from $X$ to $Y$. Then
a linear operator $V$ from $X$ to $Y$ is in $\mathcal{L}(X,Y)$
if and only if, for any $p\in P$, there exist $q\in Q$ and $C>0$
such that\[
|Vx|_{Y,p}\leq C|x|_{X,q},\quad x\in X.\]

\end{lem}
Now we define an integral kernel operator.

\begin{prop}
[\bf Integral kernel operator]\label{thm:integral_kernel_op}Let
$\kappa\in((E^{\wedge2})^{\otimes(l+m)})^{*}$. For $\phi:=\sum_{n=0}^{\infty}\phi_{n}\in\mathcal{E}_{+}$,
$\phi_{n}\in E^{\wedge(2n)}$, let\[
\Xi_{l,m}(\kappa)\phi:=\sum_{n=0}^{\infty}\frac{(2n+2m)!}{(2n)!}\kappa\wedge_{2m}\phi_{m+n}.\]
Then \begin{equation}
\left\Vert \Xi_{l,m}(\kappa)\phi\right\Vert _{p}\leq\rho^{-\frac{r}{2}}((2l)^{2l}(2m)^{2m})^{\frac{1}{2}}\left(\frac{\rho^{-\frac{r}{2}}}{-re\log\rho}\right)^{l+m}|\kappa|_{2l,2m;p,-q}\left\Vert \phi\right\Vert _{\mathrm{max}\{ p,q\}+r}\label{eq:estimation for integral kernel operator}\end{equation}
for $\phi\in\mathcal{E}_{+}$, $p$, $q\in\mathbf{R}$, and $r>0$.
That is, $\Xi_{l,m}(\kappa)\in\mathcal{L}(\mathcal{E}_{+},\mathcal{E}_{+}^{*})$.
We call $\Xi_{l,m}(\kappa)$ an integral kernel operator with a kernel
distribution $\kappa$.
\end{prop}
\begin{proof}
Let $p\in\mathbf{R}$, $q\in\mathbf{R}$, and $r>0$. We have\begin{align*}
\|\Xi_{l,m} & (\kappa)\phi\|_{p}^{2}\\
 & =\sum_{n=0}^{\infty}\left(\frac{(2n+2m)!}{(2n)!}\right)^{2}(2n+2l)!|\kappa\wedge_{2m}\phi_{m+n}|_{p}^{2}\\
 & \leq\sum_{n=0}^{\infty}\left(\frac{(2n+2m)!}{(2n)!}\right)^{2}(2n+2l)!\rho^{4rn}|\kappa|_{2l,2m;p,-q}^{2}|\phi_{m+n}|_{\mathrm{max}\{ p,q\}+r}^{2}\\
 & =|\kappa|_{2l,2m;p,-q}^{2}\sum_{n=0}^{\infty}\frac{(2n+2m)!}{(2n)!}\frac{(2n+2l)!}{(2n)!}\rho^{4rn}\cdot(2n+2m)!|\phi_{m+n}|_{\mathrm{max}\{ p,q\}+r}^{2}\\
 & \leq M^{2}|\kappa|_{2l,2m;p,-q}^{2}\left\Vert \phi\right\Vert _{\mathrm{max}\{ p,q\}+r}^{2},\end{align*}
where\begin{align*}
M & :=\sup_{n\geq0}\,\left(\frac{(2n+2l)!}{(2n)!}\right)^{\frac{1}{2}}\left(\frac{(2n+2m)!}{(2n)!}\right)^{\frac{1}{2}}\rho^{2rn}\\
 & \leq\left(\sup_{n\geq0}\,(2n+2l)\ldots(2n+1)\rho^{2rn}\right)^{\frac{1}{2}}\left(\sup_{n\geq0}\,(2n+2m)\ldots(2n+1)\rho^{2rn}\right)^{\frac{1}{2}}.\end{align*}
Since\begin{equation}
\sup_{x\geq0}\,(x+m)\ldots(x+1)\rho^{cx}\leq\rho^{-\frac{c}{2}}m^{m}\left(\frac{\rho^{-\frac{c}{2}}}{-ce\log\rho}\right)^{m}\label{eq:Lemma 416 of Obata book 1994}\end{equation}
for any $c>0$, $m\in\mathbf{N}$ (See Lemma 4.1.6 of \cite{Obata-book-1994}),
we have\[
M\leq\rho^{-\frac{r}{2}}((2l)^{2l}(2m)^{2m})^{\frac{1}{2}}\left(\frac{\rho^{-\frac{r}{2}}}{-re\log\rho}\right)^{l+m}.\]
Therefore \eqref{eq:estimation for integral kernel operator} holds.
\end{proof}
Note that the following map\[
(E^{\otimes2(l+m)})^{*}\ni\kappa\mapsto\Xi_{l,m}(\kappa)\in\mathcal{L}(\mathcal{E}_{+},\mathcal{E}_{+}^{*})\]
is not injective. We define\[
\mathcal{A}_{l,m}(\kappa):=\frac{1}{l!m!}\sum_{\sigma=(\sigma_{1},\sigma_{2})\in\mathfrak{S}_{l}\times\mathfrak{S}_{m}}\mathrm{sign}(\sigma_{1})\mathrm{sign}(\sigma_{2})\kappa^{\sigma},\]
where $\kappa^{\sigma}$ is defined in definition \ref{thm:def_of_alternizer}
(3). Put\[
(E^{\otimes(l+m)})_{\mathrm{alt}(l,m)}^{*}:=\{\kappa\in(E^{\otimes(l+m)})^{*}\,|\,\mathcal{A}_{l,m}(\kappa)=\kappa\,\}.\]
({}``alt'' stands for {}``alternative''.)

\begin{lem}
The map\[
((E^{\wedge2})^{\otimes(l+m)})_{\mathrm{alt}(2l,2m)}^{*}\ni\kappa\mapsto\Xi_{l,m}(\kappa)\in\mathcal{L}(\mathcal{E}_{+},\mathcal{E}_{+}^{*})\]
is injective. Moreover, for $\kappa\in((E^{\wedge2})^{\otimes(l+m)})^{*}$
and $\kappa'\in((E^{\wedge2})^{\otimes(l'+m')})$, if $\Xi_{l,m}(\kappa)=\Xi_{l',m'}(\kappa')?$,
then $l=l'$, $m=m'$, and $\mathcal{A}_{2l,2m}(\kappa)=\mathcal{A}_{2l,2m}(\kappa')$.
\end{lem}
\begin{proof}
See proposition 4.3.6 of \cite{Obata-book-1994}.
\end{proof}
We also note the following corollary of Proposition \ref{thm:integral_kernel_op}.

\begin{cor}
Let $\kappa\in((E^{\wedge2})^{\otimes(l+m)})^{*}$. Then $\Xi_{l,m}(\kappa)\in\mathcal{L}(\mathcal{E}_{+},\Gamma^{+}(H))$
if and only if $\kappa$ is in $H^{\wedge(2l)}\otimes(E^{\wedge(2m)})^{*}$.
In other words, $\Xi_{l,m}(\kappa)\in\mathcal{L}(\mathcal{E}_{+},\mathcal{E}_{+}^{*})$
is extended to an element of $\mathcal{L}(\Gamma^{+}(H),\mathcal{E}_{+}^{*})$
if and only if $\kappa$ is in $(E^{\wedge(2l)})^{*}\otimes H^{\wedge(2m)}$.
\end{cor}
\begin{proof}
Let $\kappa$ be in $H^{\wedge(2l)}\otimes(E^{\wedge(2m)})^{*}$.
Then $\Xi_{l,m}(\kappa)\in\mathcal{L}(\mathcal{E}_{+},\Gamma^{+}(H))$
follows from \eqref{eq:estimation for integral kernel operator}.

Conversely, let $\Xi_{l,m}(\kappa)\in\mathcal{L}(\mathcal{E}_{+},\Gamma^{+}(H))$.
From the definition of an integral kernel operator, we have\begin{equation}
\langle\!\langle\Xi_{l,m}(\kappa)\psi_{m},\phi_{l}\rangle\!\rangle=(2l)!(2m)!\langle\kappa,\phi_{l}\otimes\psi_{m}\rangle\label{eq: relation Integral kernel op and kernel function}\end{equation}
for all $\phi_{l}\in E^{\wedge(2l)}$, $\psi_{m}\in E^{\wedge(2m)}$.
Due to continuity of $\Xi_{l,m}(\kappa)$, the left hand side of \eqref{eq: relation Integral kernel op and kernel function}
is defined for $\phi_{l}\in H^{\wedge(2l)}$. Therefore $\kappa$
is in $H^{\wedge(2l)}\otimes(E^{\wedge(2m)})^{*}$. 
\end{proof}
We are now able to see the Fock expansion for the even part of the
Fermion system. 

\begin{thm}
[\bf Fock expansion]\label{thm:operator_Fock_expansion_even_part}
For any $\Xi\in\mathcal{L}(\mathcal{E}_{+},\mathcal{E}_{+}^{*})$,
there exists a unique $\{\kappa_{l,m}\}_{l,m=0}^{\infty}$, $\kappa_{l,m}\in((E^{\wedge2})^{\otimes(l+m)})_{\mathrm{alt}(2l,2m)}^{*}$
such that \begin{equation}
\Xi\phi=\sum_{l,m=0}^{\infty}\Xi_{l,m}(\kappa_{l,m})\phi,\quad\phi\in\mathcal{E}_{+}\label{eq:Fock_expansion}\end{equation}
where the right hand side of \eqref{eq:Fock_expansion} converges in
$\mathcal{E}_{+}^{*}$. 

If $\Xi\in\mathcal{L}(\mathcal{E}_{+},\mathcal{E}_{+})$, then\[
\kappa_{l,m}\in E^{\wedge(2l)}\otimes\left(E^{\wedge(2m)}\right)^{*},\quad l,m\geq0\]
and the right hand side of \eqref{eq:Fock_expansion} converges in $\mathcal{E}_{+}$.
\end{thm}
We prove Theorem \ref{thm:operator_Fock_expansion_even_part} in this
section and the following section. First, we prove algebraic part
of Theorem \ref{thm:operator_Fock_expansion_even_part}. We define
a contraction operator.

\begin{lem}
Let\[
c(l,m;n)(x):=\sum_{\mathbf{j},\mathbf{k}}\left(\sum_{\mathbf{i}}\langle x,(e(\mathbf{i})\otimes e(\mathbf{j}))\otimes(e(\mathbf{i})\otimes e(\mathbf{k}))\rangle\right)e(\mathbf{j})\otimes e(\mathbf{k}),\]
where $x\in E^{\otimes(l+m)}$ and $\mathbf{i}$, $\mathbf{j}$, $\mathbf{k}$
run the whole $\mathbf{N}^{n}$, $\mathbf{N}^{l-n}$, $\mathbf{N}^{m-n}$
respectively. We call\[
c(l,m;n)\in\mathcal{L}(E^{\otimes(l+m)},E^{\otimes(l+m-2n)})\]
a contraction operator. Contraction operators satisfy the following
relation.\[
c(l-n_{1},m-n_{1};n_{2})c(l,m;n_{1})=c(l,m;n_{1}+n_{2}).\]
Moreover, \[
c(l,m;n)^{*}|_{E^{\otimes(l-n)}\otimes(E^{\otimes(m-n)})^{*}}\in\mathcal{L}(E^{\otimes(l-n)}\otimes(E^{\otimes(m-n)})^{*},E^{\otimes l}\otimes(E^{\otimes m})^{*}).\]
 
\end{lem}
\begin{proof}
Since\begin{align*}
|\langle x,e(\mathbf{i})\otimes e(\mathbf{j})\otimes e(\mathbf{i})\otimes e(\mathbf{k})\rangle| & \leq|x|_{p+\alpha}|e(\mathbf{i})\otimes e(\mathbf{j})\otimes e(\mathbf{i})\otimes e(\mathbf{k})|_{-(p+\alpha)}\\
 & \leq|x|_{p+\alpha}|e(\mathbf{i})|_{-\alpha}|e(\mathbf{j})|_{-(p+\alpha)}|e(\mathbf{k})|_{-(q+\alpha)}\end{align*}
for $x\in E^{\otimes(l+m)}$, $p\geq0$, $\mathbf{i}\in\mathbf{N}^{n}$,
$\mathbf{j}\in\mathbf{N}^{l-n}$, and $\mathbf{k}\in\mathbf{N}^{m-n}$,
we have\begin{align*}
|c(l,m;n)(x)|_{l,m;p,q}^{2} & =\sum_{\mathbf{j},\mathbf{k}}\left|\sum_{\mathbf{i}}\langle x,e(\mathbf{i})\otimes e(\mathbf{j})\otimes e(\mathbf{i})\otimes e(\mathbf{k})\rangle\right|^{2}|e(\mathbf{j})|_{p}^{2}|e(\mathbf{k})|_{q}^{2}\\
 & \leq|x|_{p+\alpha}^{2}\sum_{\mathbf{j},\mathbf{k}}\sum_{\mathbf{i}}|e(\mathbf{i})|_{-\alpha}^{2}|e(\mathbf{j})|_{-\alpha}^{2}|e(\mathbf{k})|_{-\alpha}^{2}\\
 & =\delta^{2(l-n)+2(m-n)+2n}|x|_{p+\alpha}^{2}.\end{align*}
This implies continuity of the linear operator $c(l,m;n)$ from $E^{\otimes(l+m)}$
to $E^{\otimes(l+m-2n)}$. 

Next, we consider $c(l,m;n)^{*}$. For each $p>0$, there exists $q\in\mathbf{R}$
such that $|x|_{l-n,m-n;p,-\max\{ p+\alpha,q\}-\alpha}$ is finite.
Then\begin{align*}
|c & (l,m;n)^{*}(x)|_{l,m;p,-\max\{ p+\alpha,q\}-\alpha}^{2}\\
 & =\sum_{\mathbf{i},\mathbf{j},\mathbf{k},\mathbf{k}'}|\langle x,c(l,m;n)(e(\mathbf{i})\otimes e(\mathbf{j})\otimes e(\mathbf{i}')\otimes e(\mathbf{k})\rangle|^{2}\\
 & \qquad\qquad\times|e(\mathbf{i})\otimes e(\mathbf{j})\otimes e(\mathbf{i}')\otimes e(\mathbf{k})|_{l,m;p,-\max\{ p+\alpha,q\}-\alpha}^{2}\\
 & =|x|_{l-n,m-n;p,-\max\{ p+\alpha,q\}-\alpha}^{2}\sum_{\mathbf{k},\mathbf{k}'}|\langle e(\mathbf{i}),e(\mathbf{i}')\rangle|^{2}|e(\mathbf{i})|_{p}^{2}|e(\mathbf{i}')|_{-\max\{ p+\alpha,q\}-\alpha}^{2}\\
 & \leq|x|_{l-n,m-n;p,-\max\{ p+\alpha,q\}-\alpha}^{2}\sum_{\mathbf{k},\mathbf{k}'}|e(\mathbf{i})|_{-(p+\alpha)}^{2}|e(\mathbf{i}')|_{p+\alpha}^{2}|e(\mathbf{i})|_{p}^{2}|e(\mathbf{i}')|_{-\max\{ p+\alpha,q\}-\alpha}^{2}\\
 & \leq\delta^{4n}|x|_{l-n,m-n;p,-\max\{ p+\alpha,q\}-\alpha}^{2}.\end{align*}
for all $x\in E^{\otimes(l-n)}\otimes(E^{\otimes(m-n)})^{*}$. Here
$\mathbf{i}$, $\mathbf{i}'$, $\mathbf{j}$, $\mathbf{k}$ run the
whole $\mathbf{N}^{n}$, $\mathbf{N}^{n}$, $\mathbf{N}^{l-n}$ and
$\mathbf{N}^{m-n}$ respectively. Therefore the restriction of $c^{*}(l,m;n)$
to $E^{\otimes(l-n)}\otimes(E^{\otimes(m-n)})^{*}$ is continuous
linear operator from $E^{\otimes(l-n)}\otimes(E^{\otimes(m-n)})^{*}$
to $E^{\otimes l}\otimes(E^{\otimes m})^{*}$. 
\end{proof}
\begin{defn}
Let $\Xi\in\mathcal{L}(\mathcal{E}_{+},\mathcal{E}_{+}^{*})$. We
define a continuous linear functional $\kappa_{l,m}$ on $E^{\wedge(2l)}\otimes E^{\wedge(2m)}$,
i.e., $\kappa_{l,m}\in(E^{\wedge(2l)}\otimes E^{\wedge(2m)})^{*}$,
$l,m\in\mathbf{Z}_{\geq0}$ inductively as follows :\begin{equation}
\langle\kappa_{l,m},x\rangle:=\frac{1}{(2l)!(2m)!}\langle\!\langle(\Xi\otimes1)^{*}\tau,x\rangle\!\rangle-\sum_{n=1}^{\mathrm{min}\{ l,m\}}\frac{1}{(2n)!}\langle\kappa_{l-n,m-n},c(2l,2m;2n)(x)\rangle\label{eq:definition of integral kernel function}\end{equation}
for $x\in E^{\wedge(2l)}\otimes E^{\wedge(2m)}$. Here $\tau\in\mathcal{E}_{+}\otimes\mathcal{E}_{+}^{*}$
is defined by\[
\tau(\Psi\otimes\phi)=\langle\!\langle\Psi,\phi\rangle\!\rangle,\quad\Psi\in\mathcal{E}_{+}^{*},\,\phi\in\mathcal{E}_{+}.\]
(Continuity of $\kappa_{l,m}$ is considered in the following section.)
When $x=\psi_{l}\otimes\phi_{m}\in E^{\wedge(2l)}\otimes E^{\wedge(2m)}$,
\eqref{eq:definition of integral kernel function} implies\begin{align*}
\langle\!\langle\Xi\phi_{m},\psi_{l}\rangle\!\rangle & =\sum_{n=0}^{\mathrm{min}\{ l,m\}}\frac{(2l)!(2m)!}{(2n)!}\langle\kappa_{l-n,m-n},c(2l,2m;2n)(\psi_{l}\otimes\phi_{m})\rangle\\
 & =\sum_{n=0}^{\mathrm{min}\{ l,m\}}\frac{(2l)!(2m)!}{(2n)!}\langle\kappa_{l-n,m-n},\psi_{l}\otimes^{2n}\phi_{m}\rangle\\
 & =\sum_{n=0}^{\mathrm{min}\{ l,m\}}(2l)!\left\langle \frac{(2m)!}{(2n)!}\kappa_{l-n,m-n}\wedge_{2m-2n}\phi_{m},\psi_{l}\right\rangle .\end{align*}
This shows\begin{align*}
\langle\!\langle\Xi\phi,\psi\rangle\!\rangle & =\sum_{l,m=0}^{\infty}\langle\!\langle\Xi\phi_{m},\psi_{l}\rangle\!\rangle\\
 & =\sum_{l,m=0}^{\infty}\sum_{n=0}^{\mathrm{min}\{ l,m\}}(2l)!\left\langle \frac{(2m)!}{(2n)!}\kappa_{l-n,m-n}\wedge_{2m-2n}\phi_{m},\psi_{l}\right\rangle \\
 & =\sum_{l,m=0}^{\infty}\sum_{n=0}^{\infty}(2l+2n)!\left\langle \frac{(2m+2n)!}{(2n)!}\kappa_{l,m}\wedge_{2m}\phi_{m+n},\psi_{l+n}\right\rangle \\
 & =\sum_{l,m=0}^{\infty}\langle\!\langle\Xi_{l,m}(\kappa_{l,m})\phi,\psi\rangle\!\rangle\end{align*}
for $\phi=\sum_{m}\phi_{m}$, $\psi=\sum_{l}\psi_{l}\in\mathcal{E}$,
$\phi_{m}\in E^{\wedge(2m)}$, $\psi_{l}\in E^{\wedge(2l)}$. Thus
we obtain the formal or algebraic part of Theorem \ref{thm:operator_Fock_expansion_even_part}.
It is helpful to give another algebraic expression of $\kappa_{l,m}$
before proving the analytic part of Theorem \ref{thm:operator_Fock_expansion_even_part}.
This expression is used in Lemma \ref{lem:estimation for kappa_(l,m)}.

Let\begin{equation}
K_{l,m}:=\sum_{n=0}^{\min\{ l,m\}}\frac{(2l)!(2m)!}{(2n)!}\, c(2l,2m;2n)^{*}(\kappa_{l-n,m-n})\label{eq:definition of K_(l,m)}\end{equation}
for all $l$, $m\in\mathbf{Z}_{\geq0}$. ($K_{l,m}$ is related to
the symbol of $\Xi$. The {}``symbol'' of $\Xi$is defined in Definition
\ref{def:symbol_of_operator}.)
\end{defn}
\begin{lem}
Let\[
\Pi_{k}:=\{(k_{1},\ldots,k_{t})\in\mathbf{N}^{t}\,|\, t\in\{1,2,\ldots,\min\{ l,m\}\},k_{1}+\ldots+k_{t}=k\}\]
for $k\in\mathbf{N}$. For $(k_{1},\ldots,k_{t})\in\Pi_{k}$, put\begin{gather*}
A(k_{1},\ldots,k_{t}):=\frac{1}{(2k_{1})!\ldots(2k_{t})!}\, c(2l,2m;2k)^{*}(K_{l-k,m-k}),\\
a(k_{1},\ldots,k_{t}):=\frac{1}{(2k_{1})!\ldots(2k_{t})!}\, c(2l,2m;2k)^{*}(\kappa_{l-k,m-k}).\end{gather*}

\begin{enumerate}
\item It holds that\[
a(k_{1},\ldots,k_{t})=\frac{1}{(2l-2k)!(2m-2k)!}A(k_{1},\ldots,k_{t})-\sum_{n=1}^{\min\{ l-k,m-k\}}a(k_{1},\ldots,k_{t},n).\]

\item From $(1)$, we have\begin{align*}
\kappa_{l,m} & =\frac{1}{(2l)!(2m)!}K_{l,m}\\
 & \quad+\sum_{k=1}^{\min\{ l,m\}}\frac{1}{(2l-2k)!(2m-2k)!}\left(\sum_{(k_{1},\ldots,k_{t})\in\Pi_{k}}\frac{(-1)^{t}}{(2k_{1})!\ldots(2k_{t})!}\right)\\
 & \qquad\times c(2l,2m;2k)^{*}(K_{l-k,m-k})\end{align*}
for all $l$, $m\in\mathbf{Z}_{\geq0}$ with $\min\{ l,m\}\geq1$.
\[
\kappa_{l,m}=\frac{1}{(2l)!(2m)!}K_{l,m}\]
 holds if $\min\{ l,m\}=0$.
\end{enumerate}
\end{lem}
\begin{proof}
(1) Since\begin{equation}
\kappa_{l,m}=\frac{1}{(2l)!(2m)!}\, K_{l,m}-\sum_{n=1}^{\min\{ l,m\}}\frac{1}{(2n)!}\, c(2l,2m;2n)^{*}(\kappa_{l-n,m-n})\label{eq:relation of K_(l,m) and kappa_(l,m)}\end{equation}
from the definition of $K_{l,m}$, we have

\begin{align*}
a & (k_{1},\ldots,k_{t})\\
 & =\frac{1}{(2k_{1})!\ldots(2k_{t})!}\, c(2l,2m;2k)^{*}(\kappa_{l-k,m-k})\\
 & =\frac{1}{(2k_{1})!\ldots(2k_{t})!}\, c(2l,2m;2k)^{*}\left\{ \frac{1}{(2l-2k)!(2m-2k)!}(K_{l-k,m-k})\right.\\
 & \qquad-\sum_{n=1}^{\min\{ l-k,m-k\}}\left.\frac{1}{(2n)!}\, c(2l-2k,2m-2k;2n)^{*}(\kappa_{l-k-n,m-k-n})\right\} \\
 & =\frac{1}{(2l-2k)!(2m-2k)!}\, A(k_{1},\ldots,k_{t})-\sum_{n=1}^{\min\{ l-k,m-k\}}a(k_{1},\ldots,k_{t},n).\end{align*}
(2) is proved immediately since we have \eqref{eq:relation of K_(l,m) and kappa_(l,m)}
and can show \begin{align*}
c(2l, & 2m;2k_{1})^{*}(\kappa_{l-k_{1},m-k_{1}})\\
 & =a(k_{1})\\
 & =\frac{1}{(2l-2k_{1})!(2m-2k_{1})!}\, A(k_{1})\\
 & \quad+(-1)\frac{1}{(2l-2k_{1}-2k_{2})!(2m-2k_{1}-2k_{2})!}\sum_{k_{2}=1}^{\min\{ l-k_{1},m-k_{1}\}}A(k_{1},k_{2})\\
 & \quad+\ldots\\
 & \quad(-1)^{t-1}\frac{1}{(2l-2k_{1}-\ldots-2k_{t})!(2m-2k_{1}-\ldots-2k_{t})!}\\
 & \quad\times\sum_{k_{2}=1}^{\min\{ l-k_{1},m-k_{1}\}}\ldots\sum_{k_{t}=1}^{\min\{ l-k_{1}-\ldots-k_{t-1},m-k_{1}-\ldots-k_{t-1}\}}\frac{1}{(2k_{2})!\ldots(2k_{t})!}\, A(k_{1},\ldots,k_{t})\end{align*}
by using (1) inductively.
\end{proof}

\section{The proof of the analytic part of Theorem \ref{thm:operator_Fock_expansion_even_part}}

In this section, we prove the analytic part of Theorem \ref{thm:operator_Fock_expansion_even_part}. 

\begin{defn}
\label{def:symbol_of_operator}For $\Xi\in\mathcal{L}(\mathcal{E}_{+},\mathcal{E}_{+}^{*})$,
let\[
\widehat{\Xi}(\zeta,\eta):=\left\langle \!\left\langle \Xi e^{+}(\zeta),e^{+}(\eta)\right\rangle \!\right\rangle ,\quad\zeta,\eta\in E^{\wedge2}.\]
Then we call $\widehat{\Xi}$ the symbol of $\Xi$.
\end{defn}
\begin{prop}
\label{thm:symbol_of_op_holomorphic}For any $\zeta_{1},\zeta_{2},\eta_{1},\eta_{2}\in E^{\wedge2}$,
a function\[
\mathbf{C}^{2}\ni(z,w)\mapsto\widehat{\Xi}(z\zeta_{1}+\zeta_{2},w\eta_{1}+\eta_{2})\in\mathbf{C}\]
is holomorphic in $\mathbf{C}^{2}$. 
\end{prop}
\begin{proof}
This statement follows from proposition \ref{thm:S-trans_is_holomorphic}
and\begin{align*}
\widehat{\Xi}(z\zeta_{1}+\zeta_{2},w\eta_{1}+\eta_{2}) & =S(\Xi e^{+}(z\zeta_{1}+\zeta_{2}))(w\eta_{1}+\eta_{2})\\
 & =S(\Xi^{*}e^{+}(w\eta_{1}+\eta_{2}))(z\zeta_{1}+\zeta_{2}).\end{align*}
 
\end{proof}
Moreover, symbol $\widehat{\Xi}$ satisfies the two following lemmas.

\begin{lem}
\label{thm:estimation for symbol of operator}$\,$
\begin{enumerate}
\item Let $\Xi\in\mathcal{L}(\mathcal{E}_{+},\mathcal{E}_{+}^{*})$ and
$r\geq0$. Then there exist $p\in\mathbf{R}$, $q\in\mathbf{R}$ and
$C_{0}>0$ such that\begin{equation}
|\widehat{\Xi}(\zeta,\eta)|\leq C_{0}\exp\left[\frac{\rho^{4r}}{8}(|\zeta|_{\max\{ p,q\}+r}^{2}+|\eta|_{-p}^{2})\right].\label{eq:estimation for symbol of operator}\end{equation}

\item Let $\Xi\in\mathcal{L}(\mathcal{E}_{+},\mathcal{E}_{+})$ and $r\geq0$.
Then, for any $p\geq0$, there exist $q>0$, and $C_{0}>0$ satisfying
\eqref{eq:estimation for symbol of operator}. 
\end{enumerate}
\end{lem}
\begin{proof}
We remark that\[
\| e^{+}(\zeta)\|_{\alpha}^{2}=\sum_{n=0}^{\infty}(2n)!\left|\frac{\zeta^{\wedge n}}{(2n)!}\right|_{\alpha}^{2}\leq\sum_{n=0}^{\infty}\frac{1}{n!}\left(\frac{1}{4}|\zeta|_{\alpha}^{2}\right)^{n}=\exp\left(\frac{1}{4}|\zeta|_{\alpha}^{2}\right)\]
for all $\alpha\in\mathbf{R}$ and $\zeta\in E^{\wedge2}$. 

(1) For $r\geq0$, there exists $p\leq-r$ and $C_{0}>0$ such that\[
\|\Xi e^{+}(\zeta)\|_{p+r}\leq C_{0}\| e^{+}(\zeta)\|_{-(p+r)}\leq C_{0}\exp\left(\frac{1}{8}|\zeta|_{-(p+r)}^{2}\right)\leq C_{0}\exp\left(\frac{\rho^{4r}}{8}|\zeta|_{-p}^{2}\right)\]
for all $\zeta\in E^{\wedge2}$. Thus\begin{align*}
|\widehat{\Xi}(\zeta,\eta)| & \leq\|\Xi e^{+}(\zeta)\|_{p+r}\| e^{+}(\eta)\|_{-(p+r)}\\
 & \leq C_{0}\exp\left[\frac{\rho^{4r}}{8}(|\zeta|_{-p}^{2}+|\eta|_{-p}^{2})\right]\\
 & \leq C_{0}\exp\left[\frac{\rho^{4r}}{8}(|\zeta|_{\max\{ p,-p\}+r}^{2}+|\eta|_{-p}^{2})\right].\end{align*}

(2) For any $r\geq0$ and $p\geq0$, there exist $q>0$ and $C_{0}>0$
such that\[
\|\Xi e^{+}(\zeta)\|_{p+r}\leq C_{0}\| e^{+}(\zeta)\|_{q}\leq C_{0}\exp\left(\frac{1}{8}|\zeta|_{q}^{2}\right)\leq C_{0}\exp\left(\frac{\rho^{4r}}{8}|\zeta|_{q+r}^{2}\right).\]
Hence\begin{align*}
|\widehat{\Xi}(\zeta,\eta)| & \leq\|\Xi e^{+}(\zeta)\|_{p+r}\| e^{+}(\eta)\|_{-(p+r)}\\
 & \leq C_{0}\exp\left[\frac{\rho^{4r}}{8}(|\zeta|_{q+r}^{2}+|\eta|_{-p}^{2})\right]\\
 & \leq C_{0}\exp\left[\frac{\rho^{4r}}{8}(|\zeta|_{\max\{ p,q\}+r}^{2}+|\eta|_{-p}^{2})\right].\end{align*}

\end{proof}
Now we have to remark the following lemma.

\begin{lem}
\label{thm:holomorphic_func_|a(lm)|_ineq}Let $f$ be a holomorphic
function on $\mathbf{C}$ with Taylor expansion\[
f(z,w)=\sum_{l,m=0}^{\infty}a_{l,m}z^{l}w^{m}\]
and let $f$ satisfy\[
|f(z,w)|\leq C\exp(K_{1}|z|^{2}+K_{2}|w|^{2}),\quad z,\, w\in\mathbf{C}\]
for some $C\geq0$ and $K_{1},K_{2}\geq0$. Then\[
|a_{l,m}|\leq C\left(\frac{2eK_{1}}{l}\right)^{\frac{l}{2}}\left(\frac{2eK_{2}}{m}\right)^{\frac{m}{2}}.\]

\end{lem}
\begin{proof}
See Lemma 4.4.8 of \cite{Obata-book-1994}. 
\end{proof}
\begin{lem}
\label{lem:estimation for K_(l,m)}Let $K_{l,m}$ be elements of $(E^{\wedge(2l)}\otimes E^{\wedge(2m)})^{*}$
given by \eqref{eq:definition of K_(l,m)}. Let $p$, $q\in\mathbf{R}$,
$r\geq0$, $C_{0}>0$ be numbers given in Lemma \ref{thm:estimation for symbol of operator}.
Then it holds that\begin{equation}
|\langle K_{l,m},\eta^{\wedge l}\otimes\zeta^{\wedge m}\rangle|\leq C_{0}\left(\frac{e\rho^{4r}}{4l}\right)^{\frac{l}{2}}\left(\frac{e\rho^{4r}}{4m}\right)^{\frac{m}{2}}|\eta|_{-p}^{l}|\zeta|_{\max\{ p,q\}+r}^{m}\label{eq:estimation for K_(l,m) Ver1}\end{equation}
for any $\zeta$, $\eta\in E^{\wedge2}$ and $l$, $m\geq0$. Therefore
we have\begin{equation}
|K_{l,m}|_{2l,2m;p,-\max\{ p+\alpha,q\}-r-\alpha}\leq C_{0}(e\delta^{4}\rho^{4r})^{\frac{l+m}{2}}\left(\frac{1}{(2l)!(2m)!}\right)^{\frac{1}{2}}.\label{eq:estimation for K_(l,m) Ver2}\end{equation}

\end{lem}
\begin{proof}
\eqref{eq:estimation for K_(l,m) Ver1} follows from Lemma \ref{thm:estimation for symbol of operator},
Lemma \ref{thm:holomorphic_func_|a(lm)|_ineq} and\[
\widehat{\Xi}(z\zeta,w\eta)=\sum_{l,m=0}^{\infty}\left\langle \!\!\left\langle \Xi\frac{\zeta^{\wedge m}}{(2m)!},\,\frac{\eta^{\wedge l}}{(2l)!}\right\rangle \!\!\right\rangle w^{l}z^{m}=\sum_{l,m=0}^{\infty}\langle K_{l,m},\eta^{\wedge l}\otimes\zeta^{\wedge m}\rangle w^{l}z^{m}\]
for any $z$, $w\in\mathbf{C}$. We prove \eqref{eq:estimation for K_(l,m) Ver2}.
Let $e_{i}$, $\lambda_{i}$ be C.O.N.S. and eigenvalues given in
Definition \ref{thm:property_of_self-adj_op_A} respectively. Now
fix $i_{1}<\ldots<i_{2l}$, $j_{1}<\ldots<j_{2m}$ and we put\begin{gather*}
\eta=(\lambda_{i_{1}}\ldots\lambda_{i_{2l}})^{-(p+\alpha)}\sum_{s=1}^{l}(\lambda_{i_{2s-1}}\lambda_{i_{2s}})^{p+\alpha}e_{i_{2s-1}}\wedge e_{i_{2s}},\\
\zeta=(\lambda_{j_{1}}\ldots\lambda_{j_{2m}})^{\max\{ p+\alpha,q\}+r}\sum_{t=1}^{m}(\lambda_{j_{2t-1}}\lambda_{j_{2t}})^{-(\max\{ p+\alpha,q\}+r)}e_{j_{2t-1}}\wedge e_{j_{2t}}.\end{gather*}
Then\begin{gather*}
e_{i_{1}}\wedge\ldots\wedge e_{i_{2l}}=\frac{1}{l!}(\lambda_{i_{1}}\ldots\lambda_{i_{2l}})^{(p+\alpha)(l-1)}\eta^{\wedge l},\\
e_{j_{1}}\wedge\ldots\wedge e_{j_{2m}}=\frac{1}{m!}(\lambda_{j_{1}}\ldots\lambda_{j_{2m}})^{-(\max\{ p+\alpha,q\}+r)(m-1)}\zeta^{\wedge m},\\
|\eta|_{-(p+\alpha)}^{2l}=l^{l}(\lambda_{i_{1}}\ldots\lambda_{i_{2l}})^{-2(p+\alpha)l},\\
|\zeta|_{\max\{ p+\alpha,q\}+r}^{2m}=m^{m}(\lambda_{j_{1}}\ldots\lambda_{j_{2m}})^{2(\max\{ p+\alpha,q\}+r)m}.\end{gather*}
Thus, from \eqref{eq:estimation for K_(l,m) Ver1}, we have\begin{align*}
|\langle K_{l,m} & ,(e_{i_{1}}\wedge\ldots\wedge e_{i_{2l}})\otimes(e_{j_{1}}\wedge\ldots\wedge e_{j_{2m}})\rangle|^{2}\\
 & \leq C_{0}^{2}\left(\frac{1}{l!m!}\right)^{2}\left(\frac{e\rho^{4r}}{4l}\right)^{l}\left(\frac{e\rho^{4r}}{4m}\right)^{m}l^{l}m^{m}\\
 & \quad\times(\lambda_{i_{1}}\ldots\lambda_{i_{2l}})^{-2(p+\alpha)}(\lambda_{j_{1}}\ldots\lambda_{j_{2m}})^{2(\max\{ p+\alpha,q\}+r)}\\
 & \leq C_{0}^{2}(e\rho^{4r})^{l+m}\frac{1}{(2l)!(2m)!}(\lambda_{i_{1}}\ldots\lambda_{i_{2l}})^{-2(p+\alpha)}(\lambda_{j_{1}}\ldots\lambda_{j_{2m}})^{2(\max\{ p+\alpha,q\}+r)}.\end{align*}
Here we used the following inequality \[
(2l)!=2l\cdot(2l-1)\cdot\ldots2\cdot1\leq2l\cdot2l\cdot\ldots\cdot2\cdot2=(2^{l}l!)^{2}.\]
Therefore\begin{align*}
|K_{l,m} & |_{2l,2m;p,-(\max\{ p+\alpha,q\}+r+\alpha)}^{2}\\
 & =\sum_{i_{1}<\ldots<i_{2l},j_{1}<\ldots<j_{2m}}|\langle K_{l,m},(e_{i_{1}}\wedge\ldots\wedge e_{i_{2l}})\otimes(e_{j_{1}}\wedge\ldots\wedge e_{j_{2m}})\rangle|^{2}\\
 & \qquad\qquad\times|e_{i_{1}}\wedge\ldots\wedge e_{i_{2l}}|_{p}^{2}|e_{j_{1}}\wedge\ldots\wedge e_{j_{2m}}|_{-(\max\{ p+\alpha,q\}+r+\alpha)}^{2}\\
 & \leq C_{0}^{2}(e\rho^{4r})^{l+m}\frac{1}{(2l)!(2m)!}\sum_{i_{1}<\ldots<i_{2l},j_{1}<\ldots<j_{2m}}(\lambda_{i_{1}}\ldots\lambda_{i_{2l}})^{-2\alpha}(\lambda_{j_{1}}\ldots\lambda_{j_{2m}})^{-2\alpha}\\
 & \leq C_{0}^{2}(e\delta^{4}\rho^{4r})^{l+m}\frac{1}{(2l)!(2m)!}.\end{align*}

\end{proof}
Next, we discuss estimations for $c(2l,2m;2k)^{*}(K_{l-k,m-k})$,
$k=1,2$, $\ldots$, $\min\{ l,m\}$. 

\begin{lem}
\label{thm:estimation for symbol of integral kernel op}$\,$
\begin{enumerate}
\item Let $\lambda_{l,m}\in(E^{\wedge(2l)}\otimes E^{\wedge(2m)})^{*}$
and $r\geq0$. Then there exist $p\in\mathbf{R}$ and $q\in\mathbf{R}$
with $|\lambda_{l,m}|_{2l,2m;p+r,-q}<+\infty$ and we have\begin{equation}
|\widehat{\Xi_{l,m}(\lambda_{l,m})}(\zeta,\eta)|\leq(l!m!)^{\frac{1}{2}}|\lambda_{l,m}|_{2l,2m;p+r,-q}\exp\left[\frac{\rho^{4r}}{2}(|\zeta|_{\max\{ p,q\}+2r}^{2}+|\eta|_{-p}^{2})\right].\label{eq:estimation for symbol of int ker op}\end{equation}

\item Let $\lambda_{l,m}\in E^{\wedge(2l)}\otimes(E^{\wedge(2m)})^{*}$
and $r\geq0$. Then, for any $p\geq0$, there exists $q\in\mathbf{R}$
satisfying \eqref{eq:estimation for symbol of int ker op}.
\end{enumerate}
\end{lem}
\begin{proof}
(1) From definition of an integral kernel operator, we have\begin{align*}
| & \widehat{\Xi_{l,m}(\lambda_{l,m})}(\zeta,\eta)|\\
 & =\left|\sum_{n=0}^{\infty}\frac{1}{(2n)!}\langle\lambda_{l,m}\wedge_{2m}\zeta^{\wedge(m+n)},\eta^{\wedge(l+n)}\rangle\right|\\
 & \leq\sum_{n=0}^{\infty}\frac{1}{(2n)!}|\lambda_{l,m}\wedge_{2m}\zeta^{\wedge(m+n)}|_{p+r}|\eta^{\wedge(l+n)}|_{-(p+r)}\\
 & \leq\sum_{n=0}^{\infty}\frac{1}{(2n)!}\left(\rho^{2(m+n)r}|\lambda_{l,m}|_{2l,2m;p+r,-q}|\zeta^{\wedge(m+n)}|_{\max\{ p+r,q\}+r}\right)\left(\rho^{2(l+n)r}|\eta^{\wedge(l+n)}|_{-p}\right)\\
 & \leq|\lambda_{l,m}|_{2l,2m;p+r,-q}(\rho^{4r}|\zeta|_{\max\{ p,q\}+2r}^{2})^{\frac{m}{2}}(\rho^{4r}|\eta|_{-p}^{2})^{\frac{l}{2}}\left(\sum_{n=0}^{\infty}\frac{1}{(2n)!}\rho^{4nr}|\zeta|_{\max\{ p,q\}+2r}^{n}|\eta|_{-p}^{n}\right).\end{align*}
Therefore we obtain \eqref{eq:estimation for symbol of int ker op}
by using\begin{align*}
\sum_{n=0}^{\infty}\frac{1}{(2n)!}\rho^{4nr}|\zeta|_{\max\{ p,q\}+2r}^{n}|\eta|_{-p}^{n} & \leq\sum_{n=0}^{\infty}\frac{(\rho^{2r}|\zeta|_{\max\{ p,q\}+2r})^{n}}{(n!)^{\frac{1}{2}}}\cdot\frac{(\rho^{2r}|\eta|_{-p})^{n}}{(n!)^{\frac{1}{2}}}\\
 & \leq\left\{ \sum_{n=0}^{\infty}\frac{(\rho^{4r}|\zeta|_{\max\{ p,q\}+2r}^{2})^{n}}{n!}\right\} ^{\frac{1}{2}}\left\{ \sum_{n=0}^{\infty}\frac{(\rho^{4r}|\eta|_{-p}^{2})^{n}}{n!}\right\} ^{\frac{1}{2}}\\
 & =\exp\left(\frac{\rho^{4r}}{2}|\zeta|_{\max\{ p,q\}+2r}^{2}\right)\exp\left(\frac{\rho^{4r}}{2}|\eta|_{-p}^{2}\right)\end{align*}
and\[
(\rho^{4r}|\zeta|_{\max\{ p,q\}+2r}^{2})^{m}=m!\cdot\frac{1}{m!}(\rho^{4r}|\zeta|_{\max\{ p,q\}+2r}^{2})^{m}\leq m!\exp(\rho^{4r}|\zeta|_{\max\{ p,q\}+2r}^{2}).\]
(2) is easily checked in the same manner as (1).
\end{proof}
\begin{lem}
Let $\lambda_{l,m}$ be an element of $(E^{\wedge(2l)}\otimes E^{\wedge(2m)})^{*}$
and $p$, $q\in\mathbf{R}$, $r\geq0$ be numbers given in Lemma \ref{thm:estimation for symbol of integral kernel op}.
Then it holds that\begin{align}
|\langle c & (2l+2n,2m+2n;2n)^{*}(\lambda_{l,m}),\eta^{\wedge(l+n)}\otimes\zeta^{\wedge(m+n)}\rangle|\nonumber \\
 & \leq(l!m!)^{\frac{1}{2}}|\lambda_{l,m}|_{2l,2m;p+r,-q}\left(\frac{e\rho^{4r}}{l+n}\right)^{\frac{l+n}{2}}\left(\frac{e\rho^{4r}}{m+n}\right)^{\frac{m+n}{2}}|\eta|_{-p}^{l+n}|\zeta|_{\max\{ p,q\}+2r}^{m+n}\label{eq:inequality of contraction lambda_(l,m) Ver1}\end{align}
for all $\zeta$, $\eta\in E^{\wedge2}$. 
\end{lem}
\begin{proof}
Since \[
\widehat{\Xi(\lambda_{l,m})}(z\zeta,w\eta)=\sum_{l,m=0}^{\infty}\langle c(2l+2n,2m+2n;2n)^{*}(\lambda_{l,m}),\eta^{\wedge(l+n)}\otimes\zeta^{\wedge(m+n)}\rangle w^{l+n}z^{m+n}\]
for $z$, $w\in\mathbf{C}$, Lemma \ref{thm:estimation for symbol of integral kernel op}
and Lemma \ref{thm:holomorphic_func_|a(lm)|_ineq} imply \eqref{eq:inequality of contraction lambda_(l,m) Ver1}.
\end{proof}
\begin{lem}
Let $\lambda_{l-k,m-k}$ be an element of $(E^{\wedge(2l-2k)}\otimes E^{\wedge(2m-2k)})^{*}$
and $p$, $q\in\mathbf{R}$, $r\geq0$ be numbers given in Lemma \ref{thm:estimation for symbol of integral kernel op}.
Then\begin{align}
|c & (2l,2m;2k)^{*}(\lambda_{l-k,m-k})|_{2l,2m;p,-\max\{ p+\alpha,q\}-2r-\alpha}\nonumber \\
 & \leq\left(\frac{1}{(2l)!(2m)!}\right)^{\frac{1}{2}}((l-k)!(m-k)!)^{\frac{1}{2}}(4e\delta^{4}\rho^{4r})^{\frac{l+m}{2}}|\lambda_{l-k,m-k}|_{2(l-k),2(m-k);p+r,-q}\label{eq:inequality of contraction lambda_(l,m) Ver2}\end{align}

\end{lem}
\begin{proof}
The relation \eqref{eq:inequality of contraction lambda_(l,m) Ver1}
leads us to\begin{align*}
|\langle c & (2l,2m;2k)^{*}(\lambda_{l-k,m-k}),(e_{i_{1}}\wedge\ldots\wedge e_{i_{2l}})\otimes(e_{j_{1}}\wedge\ldots\wedge e_{j_{2m}})\rangle|^{2}\\
 & \leq(l-k)!(m-k)!|\lambda_{l-k,m-k}|_{2(l-k),2(m-k);p+r,-q}^{2}\left(\frac{e\rho^{4r}}{l}\right)^{l}\left(\frac{e\rho^{4r}}{m}\right)^{m}\\
 & \qquad\times\left(\frac{1}{l!m!}\right)^{2}l^{l}m^{m}(\lambda_{i_{1}}\ldots\lambda_{i_{2l}})^{-2(p+\alpha)}(\lambda_{j_{1}}\ldots\lambda_{j_{2m}})^{2(\max\{ p+\alpha,q\}+2r)}\\
 & \leq\frac{1}{(2l)!(2m)!}\,(l-k)!(m-k)!|\lambda_{l-k,m-k}|_{2(l-k),2(m-k);p+r,-q}^{2}(4e\rho^{4r})^{l+m}\\
 & \qquad\times(\lambda_{i_{1}}\ldots\lambda_{i_{2l}})^{-2(p+\alpha)}(\lambda_{j_{1}}\ldots\lambda_{j_{2m}})^{2(\max\{ p+\alpha,q\}+2r)}.\end{align*}
 (See Lemma \ref{lem:estimation for K_(l,m)}.) Thus we can show \eqref{eq:inequality of contraction lambda_(l,m) Ver2}
immediately.
\end{proof}
We are now ready to give the estimation for $\kappa_{l,m}$. Recall
that $\kappa_{l,m}$ is determined by $\Xi\in\mathcal{L}(\mathcal{E}_{+},\mathcal{E}_{+}^{*})$
via \eqref{eq:definition of integral kernel function} 

\begin{lem}
\label{lem:estimation for kappa_(l,m)}$\,$
\begin{enumerate}
\item Let $\Xi\in\mathcal{L}(\mathcal{E}_{+},\mathcal{E}_{+}^{*})$ and
$r\geq0$. Then there exist $p\in\mathbf{R}$, $q\in\mathbf{R}$,
$C_{1}>0$ and $C_{2}>0$ such that\begin{equation}
|\kappa_{l,m}|_{2l,2m;p,-\max\{ p+r+\alpha,q\}-3r-2\alpha}\leq C_{1}(C_{2}\rho^{2r})^{l+m}\left(\frac{1}{(2l)!(2m)!}\right)^{\frac{1}{2}}\label{eq:estimation for kappa_(l,m)}\end{equation}

\item Let $\Xi\in\mathcal{L}(\mathcal{E}_{+},\mathcal{E}_{+})$ and $r>0$.
Then, for any $p\geq0$, there exist $q>0$, $C_{1}>0$, and $C_{2}>0$
satisfying \eqref{eq:estimation for kappa_(l,m)}. 
\end{enumerate}
\end{lem}
\begin{proof}
We see only (1). From Lemma \ref{lem:estimation for K_(l,m)}, we
can take $p\in\mathbf{R}$, $q\in\mathbf{R}$, and $C_{0}>0$ satisfying\[
|K_{l-k,m-k}|_{2(l-k),2(m-k);p+r,-Q}\leq C_{0}(e\delta^{4}\rho^{4r})^{\frac{l+m-2k}{2}}\left(\frac{1}{(2l-2k)!(2m-2k)!}\right)^{\frac{1}{2}}\]
for all $l$, $m\geq0$, $0\leq k\leq\min\{ l,m\}$, where \[
Q:=\max\{ p+r+\alpha,q\}+r+\alpha.\]
Hence \eqref{eq:inequality of contraction lambda_(l,m) Ver2} implies\begin{align*}
|c & (2l,2m;2k)^{*}(K_{l-k,m-k})|_{2l,2m;p,-\max\{ p+\alpha,Q\}-2r-\alpha}\\
 & \leq\left(\frac{1}{(2l)!(2m)!}\right)^{\frac{1}{2}}((l-k)!(m-k)!)^{\frac{1}{2}}(4e\delta^{4}\rho^{4r})^{\frac{l+m}{2}}|K_{l-k,m-k}|_{2(l-k),2(m-k);p+r,-Q}.\\
 & \leq C_{0}\left(\frac{1}{(2l)!(2m)!}\right)^{\frac{1}{2}}(4e\delta^{4}\rho^{4r})^{\frac{l+m}{2}}(e\delta^{4})^{\frac{l+m-2k}{2}}\end{align*}
 Here\[
\max\{ p+\alpha,Q\}+2r+\alpha=Q+2r+\alpha=\max\{ p+r+\alpha,q\}+3r+2\alpha\]
and we obtain the following estimation for $\kappa_{l,m}$:\begin{align*}
|\kappa_{l,m} & |_{2l,2m;p,-\max\{ p+r+\alpha,q\}-3r-2\alpha}\\
 & \leq C_{0}\left(\frac{1}{(2l)!(2m)!}\right)^{\frac{1}{2}}\rho^{2r(l+m)}\left\{ \frac{C_{3}^{l+m}}{(2l)!(2m)!}\right.\\
 & \quad+\left.(2C_{3})^{l+m}\sum_{k=1}^{\min\{ l,m\}}\frac{C_{3}^{(l-k)+(m-k)}}{(2l-2k)!(2m-2k)!}\sum_{(k_{1},\ldots,k_{t})\in\Pi_{k}}\frac{1}{(2k_{1})!\ldots(2k_{t})!}\right\} \end{align*}
where $C_{3}:=(e\delta^{4})^{\frac{1}{2}}$. Since\begin{align*}
\sum_{(k_{1},\ldots,k_{t})\in\Pi_{k}}\frac{1}{(2k_{1})!\ldots(2k_{t})!} & \leq\frac{1}{k!}\sum_{(k_{1},\ldots,k_{t})\in\Pi_{k}}\frac{k!}{k_{1}!\ldots k_{t}!}=\frac{1}{k!}\sum_{t=1}^{k}(\overbrace{1+\ldots+1}^{t})^{k}\\
 & =\frac{1}{k!}\sum_{t=1}^{k}t^{k}\leq\frac{1}{k!}\int_{1}^{k+1}t^{k}\mathrm{d}t=\frac{(k+1)^{k+1}}{(k+1)!}\\
 & \leq e^{k+1},\end{align*}
it holds that\begin{align*}
\sum_{k=1}^{\min\{ l,m\}} & \frac{C_{3}^{(l-k)+(m-k)}}{(2l-2k)!(2m-2k)!}\sum_{(k_{1},\ldots,k_{t})\in\Pi_{k}}\frac{1}{(2k_{1})!\ldots(2k_{t})!}\\
 & \leq\left(\sum_{k=1}^{\min\{ l,m\}}\frac{(\sqrt{C_{3}})^{2l-2k}}{(2l-2k)!}\right)\left(\sum_{k=1}^{\min\{ l,m\}}\frac{(\sqrt{C_{3}})^{2m-2k}}{(2m-2k)!}\right)e^{\min\{ l,m\}+1}\\
 & \leq e^{2\sqrt{C_{3}}}e^{\min\{ l,m\}+1}.\end{align*}
Therefore we obtain\begin{align*}
| & \kappa_{l,m}|_{2l,2m;p,-\max\{ p+r+\alpha,q\}-3r-2\alpha}\\
 & \leq C_{0}\left(\frac{1}{(2l)!(2m)!}\right)^{\frac{1}{2}}\rho^{2r(l+m)}e^{1+2\sqrt{C_{3}}}(1+e^{\min\{ l,m\}}(2C_{3})^{l+m})\\
 & \leq C_{0}\left(\frac{1}{(2l)!(2m)!}\right)^{\frac{1}{2}}\rho^{2r(l+m)}e^{1+2\sqrt{C_{3}}}\cdot2\max\{1,2eC_{3}\}^{l+m}\\
 & =C_{1}(C_{2}\rho^{2r})^{l+m}\left(\frac{1}{(2l)!(2m)!}\right)^{\frac{1}{2}},\end{align*}
where\[
C_{1}:=2C_{0}e^{1+2\sqrt{C_{3}}},\quad C_{2}:=\max\{1,2eC_{3}\}.\]

\end{proof}
\begin{lem}
Fock expansion \eqref{eq:Fock_expansion} converges in $\mathcal{E}_{+}^{*}$(resp.
$\mathcal{E}_{+}$) with respect to the topology of $\mathcal{E}_{+}^{*}$(resp.
$\mathcal{E}_{+}$) if $\Xi\in\mathcal{L}(\mathcal{E}_{+},\mathcal{E}_{+}^{*})$
(resp. $\Xi\in\mathcal{L}(\mathcal{E}_{+},\mathcal{E}_{+})$). 
\end{lem}
\begin{proof}
Due to \eqref{eq:estimation for integral kernel operator} and \eqref{eq:estimation for kappa_(l,m)},
\begin{align*}
\| & \Xi_{l,m}(\kappa_{l,m})\phi\|_{p}\\
 & \leq\rho^{-\frac{r}{2}}(2m)^{m}(2l)^{l}\left(\frac{\rho^{-\frac{r}{2}}}{-re\log\rho}\right)^{l+m}\\
 & \qquad\times|\kappa_{l,m}|_{2l,2m;p,-\max\{ p+r+\alpha,q\}-3r-2\alpha}\|\phi\|_{\max\{ p,\max\{ p+r+\alpha,q\}+3r+2\alpha\}+r}\\
 & \leq C_{1}\rho^{-\frac{r}{2}}\left(\frac{(2m)^{2m}(2l)^{2l}}{(2l)!(2m)!}\right)^{\frac{1}{2}}\left(C_{2}\rho^{2r}\frac{\rho^{-\frac{r}{2}}}{-re\log\rho}\right)^{l+m}\|\phi\|_{\max\{ p+\alpha,q\}+4r+2\alpha}.\end{align*}
Put\[
R:=\frac{C_{2}}{2}\rho^{2r}\frac{\rho^{-\frac{r}{2}}}{\log\rho^{-\frac{r}{2}}}>0\]
By using\[
\frac{(2l)^{2l}}{(2l)!}\leq e^{2l},\quad\frac{(2m)^{2m}}{(2m)!}\leq e^{2m},\]
we have\[
\|\Xi_{l,m}(\kappa_{l,m})\phi\|_{p}\leq C_{1}\rho^{-\frac{r}{2}}R^{l+m}\|\phi\|_{\max\{ p+\alpha,q\}+4r+2\alpha}.\]
Now, if we choose sufficiently large $r>0$, then $R<1$ holds. Therefore\[
\sum_{l,m=0}^{\infty}\|\Xi_{l,m}(\kappa_{l,m})\phi\|_{p}\leq\frac{C_{1}\rho^{-\frac{r}{2}}}{(1-R)^{2}}\|\phi\|_{\max\{ p+\alpha,q\}+4r+2\alpha}\]
and this implies $\sum_{l,m=0}^{\infty}\Xi_{l,m}(\kappa_{l,m})\phi$
converges in $\mathcal{E}_{+}^{*}$ (resp. $\mathcal{E}_{+}$) with
respect to the topology of $\mathcal{E}_{+}^{*}$(resp. $\mathcal{E}_{+}$).
\end{proof}

\section{Fock expansion for the fermion system}

In this section, we extend result of section \ref{sec:Fock expansion (even)}
to the whole of the Fermion system. In order to make the white noise
calculus for the Fermion system, we mention properties for creation
and annihilation operators for the Fermion system.

\begin{defn}
$\,$
\begin{enumerate}
\item For $f\in E^{*}$, we define an annihilation operator $a(f)\in\mathcal{L}(\mathcal{E},\mathcal{E})$
as follows :\begin{gather*}
a(f):\mathcal{E}\ni\phi=(\phi_{n})_{n\in\mathbf{Z}\geq0}\mapsto a(f)\phi\in\mathcal{E},\\
a(f)\phi_{n}:=n\cdot f\wedge_{1}\phi_{n},\quad n\geq1,\\
a(f)\phi_{0}=0.\end{gather*}
(Well-definedness and continuity of $a(f)$ is discussed in the following
lemma. )
\item For $f\in E^{*}$, we define a creation operator $a^{\dagger}(f)\in\mathcal{L}(\mathcal{E}^{*},\mathcal{E}^{*})$
as follows :\begin{gather*}
a^{\dagger}(f):\mathcal{E}^{*}\ni\phi=(\phi_{n})_{n\in\mathbf{Z}\geq0}\mapsto a^{\dagger}(f)\phi\in\mathcal{E}^{*},\\
a^{\dagger}(f)\phi_{n}:=f\wedge\phi_{n},\quad n\geq0.\end{gather*}

\item For $f\in E^{*}$ and $(l,m)\in\{(1,0),(0,1)\}$, put\[
a_{(l,m)}(f):=\left\{ \begin{array}{ll}
a^{\dagger}(f), & \mathrm{if}\,(l,m)=(1,0),\\
a(f), & \mathrm{if}\,(l,m)=(0,1).\end{array}\right.\]
 
\end{enumerate}
\end{defn}
$a(f)$ is a map from $\mathcal{E}$ to $\mathcal{E}$ and is a continuous
map as follows:

\begin{lem}
Let $p$, $q\in\mathbf{R}$, $r>0$, and $f\in E^{*}$. Then\begin{equation}
\left\Vert a_{(l,m)}(f)\phi\right\Vert _{p}\leq\left(\frac{\rho^{-2r}}{-2re\log\rho}\right)^{\frac{1}{2}}|f|_{m,l;-(q+r),p}\left\Vert \phi\right\Vert _{\max\{ p,q\}+r},\quad\phi\in\mathcal{E}.\label{eq:estimation for annihilation creation op}\end{equation}
Thus we have the following properties (1)--(3). Let $\sigma$ be $+$,
$-$, or a blank.
\begin{enumerate}
\item $a(f)|_{\mathcal{E}_{\sigma}}\in\mathcal{L}(\mathcal{E}_{\sigma},\mathcal{E}_{-\sigma})$
for $f\in E^{*}$,
\item $a^{\dagger}(f)|_{\mathcal{E}_{\sigma}}\in\mathcal{L}(\mathcal{E}_{\sigma},\mathcal{E}_{-\sigma})$
for $f\in E$,
\item $a^{\dagger}(f)|_{\mathcal{E}_{\sigma}^{*}}=a(f)^{*}|_{\mathcal{E}_{\sigma}^{*}}\in\mathcal{L}(\mathcal{E}_{\sigma}^{*},\mathcal{E}_{-\sigma}^{*})$
for $f\in E^{*}$.
\end{enumerate}
\end{lem}
\begin{proof}
\eqref{eq:estimation for annihilation creation op} can be shown by
using Lemma \ref{lem:estimation for contraction (wedge)} and \eqref{eq:Lemma 416 of Obata book 1994}.
In fact, we have\begin{align*}
\left\Vert a(f)\phi\right\Vert _{p}^{2} & =\sum_{n=1}^{\infty}(n-1)!|nf\wedge^{1}\phi_{n}|_{p}^{2}\\
 & \leq\sum_{n=1}^{\infty}n!\cdot n\rho^{2nr}|f|_{-q}^{2}|\phi_{n}|_{\max\{ p,q\}+r}^{2}\\
 & \leq\left(\sup_{n\geq1}(n+1)\rho^{2nr}\right)|f|_{-q}^{2}\|\phi\|_{\max\{ p,q\}+r}^{2}\\
 & \leq\left(\frac{\rho^{-2r}}{-2re\log\rho}\right)|f|_{-q}^{2}\|\phi\|_{\max\{ p,q\}+r}^{2}\end{align*}
 and also have\begin{align*}
\left\Vert a^{\dagger}(f)\phi\right\Vert _{p}^{2} & =\sum_{n=0}^{\infty}(n+1)!|f\wedge\phi_{n}|_{p}^{2}\\
 & \leq\sum_{n=0}^{\infty}n!\cdot(n+1)\rho^{2nr}|f|_{p}^{2}|\phi_{n}|_{\max\{ p,q\}+r}^{2}\\
 & \leq\left(\sup_{n\geq1}(n+1)\rho^{2nr}\right)|f|_{p}^{2}\|\phi\|_{\max\{ p,q\}+r}^{2}\\
 & \leq\left(\frac{\rho^{-2r}}{-2re\log\rho}\right)|f|_{p}^{2}\|\phi\|_{\max\{ p,q\}+r}^{2}.\end{align*}
(1) We take any $p>0$ and $r>0$. Then we choose $q>0$ with $|f|_{-q}<+\infty$
and this implies that $a(f)|_{\mathcal{E}_{\sigma}}\in\mathcal{L}(\mathcal{E}_{\sigma},\mathcal{E}_{-\sigma})$.
(2) is obvious. (3) follows from (1).
\end{proof}
Creation and annihilation operators satisfy the following commutation
relation, called canonical anti-commutation relations. 

\begin{prop}
For $f\in E$ and $g\in E^{*}$, we have\[
\{ a^{\dagger}(f),a(g)\}\phi=\left\langle g,f\right\rangle \phi\]
for all $\phi\in\mathcal{E}_{\sigma}$. ($\sigma$ is $+$,$-$, or
a blank. ), Moreover\[
(a^{\dagger}(f)+a(Jf))^{2}\phi=\phi\]
for all $\phi\in$on $\mathcal{E}_{\sigma}$ for $f\in E$ with $(f,f)_{0}=1$.
Here $Jf\in E$ is the complex conjugate of $f\in E$.
\end{prop}
Put \[
d\Gamma(A)^{(n)}:=\sum_{i=1}^{n}1^{\otimes(i-1)}\otimes A\otimes1^{\otimes(n-i)}\]
 for a linear operator $A\in\mathcal{L}(E,E^{*})$. 

\begin{lem}
(1) For $f$, $g\in E^{*}$, we have\begin{gather}
a^{\dagger}(f)a^{\dagger}(g)|_{\mathcal{E}_{+}}=\Xi_{1,0}(f\wedge g),\label{eq:creation_creation_equal_2particle_creation}\\
a(f)a(g)|_{\mathcal{E}_{+}}=\Xi_{0,1}(f\wedge g).\label{eq:annihilation_annihilation_equal_2particle_annihilation}\end{gather}
(2) Let $(f\otimes g)h:=\left\langle g,h\right\rangle f$ for $f$,
$g\in E^{*}$, and $h\in E$. Then we have\begin{equation}
a^{\dagger}(f)a(g)|_{\mathcal{E}_{+}}=\Xi_{1,1}((1_{2}\otimes d\Gamma(f\otimes g)^{(2)})^{*}\tau)\label{eq:creation_annihilation_equal_second_quantization}\end{equation}
where $\tau\in(E^{\wedge2})^{*}\otimes(E^{\wedge2})^{*}$ is defined
by $\tau(\zeta,\eta):=\left\langle \zeta,\eta\right\rangle $ for
all $\zeta\in(E^{\wedge2})^{*}$ and $\eta\in E^{\wedge2}$. 
\end{lem}
\begin{proof}
(1) is easily checked. We show only (2). For any $h_{i}\in E$ $(i=1,2,\ldots,2n)$,
we have\begin{eqnarray*}
a^{\dagger}(f)a(g)h_{1}\wedge\ldots\wedge h_{2n} & = & \sum_{i=1}^{2n}(-1)^{i-1}\left\langle g,h_{i}\right\rangle f\wedge h_{1}\wedge\ldots\wedge h_{i-1}\wedge h_{i+1}\wedge\ldots\wedge h_{2n}\\
 & = & \sum_{i=1}^{2n}(-1)^{i-1}(f\otimes g)h_{i}\wedge h_{1}\wedge\ldots\wedge h_{i-1}\wedge h_{i+1}\wedge\ldots\wedge h_{2n}\\
 & = & d\Gamma(f\otimes g)^{(2n)}h_{1}\wedge\ldots\wedge h_{2n}\\
 & = & d\Gamma(d\Gamma(f\otimes g)^{(2)})^{(n)}h_{1}\wedge\ldots\wedge h_{2n}.\end{eqnarray*}
 Thus we have \eqref{eq:creation_annihilation_equal_second_quantization}.
(See proposition 4.6.13 of \cite{Obata-book-1994}.)
\end{proof}
Let $\Xi_{l,m}(\kappa)\in\mathcal{L}(\mathcal{E}_{+},\mathcal{E}_{+}^{*})$
be an integral kernel operator with a kernel distribution $\kappa$.
Let $f$ be an element of $E$ with $(f,f)_{0}=1$ and $W(f):=a^{\dagger}(f)+a(Jf)$.
Then we also call all operators\[
\Xi_{l,m}(\kappa)W(f),\quad W(f)^{*}\Xi_{l,m}(\kappa),\quad W(f)^{*}\Xi_{l,m}(\kappa)W(f).\]
integral kernel operators for the sake of convenience. Now we give
the main theorem of this paper.

\begin{thm}
Every $\Xi\in\mathcal{L}(\mathcal{E},\mathcal{E}^{*})$ is realized
as a series of integral kernel operators.
\end{thm}
\begin{proof}
Note that for $\Xi\in\mathcal{L}(\mathcal{E},\mathcal{E}^{*})$ there
exist unique $\Xi_{\alpha\beta}\in\mathcal{L}(\mathcal{E}_{\beta},\mathcal{E}_{\alpha}^{*})$
($\alpha,\,\beta\in\{+,-\}$) such that\[
\Xi=\Xi_{++}+\Xi_{+-}+\Xi_{-+}+\Xi_{--}.\]
Thus we have only to show that each $\Xi_{\alpha\beta}\in\mathcal{L}(\mathcal{E}_{\beta},\mathcal{E}_{\alpha}^{*})$
is realized as a series of integral kernel operators.

Let $\Xi_{+-}$ be an element of $\mathcal{L}(\mathcal{E}_{-},\mathcal{E}_{+}^{*})$
and $f\in E$ satisfy $(f,f)_{0}=1$. Note that $\Xi_{+-}W(f)$ is
an element of $\mathcal{L}(\mathcal{E}_{+},\mathcal{E}_{+}^{*})$.
Then, from theorem \ref{thm:operator_Fock_expansion_even_part}, there
exists a unique kernel distribution $\kappa_{+-}(l,m;f)$ such that\[
\Xi_{+-}W(f)=\sum_{l,m=0}^{\infty}\Xi_{l,m}(\kappa_{+-}(l,m;f)).\]
Thus we have\[
\Xi_{+-}=\Xi_{+-}W(f)^{2}=\sum_{l,m=0}^{\infty}\Xi_{l,m}(\kappa_{+-}(l,m;f))W(f).\]
 In the same manner, for $\Xi_{-+}\in\mathcal{L}(\mathcal{E}_{+},\mathcal{E}_{-}^{*})$,
we have $\kappa_{-+}(l,m;f)$ with\[
\Xi_{-+}=\left(W(f)^{*}\right)^{2}\Xi_{-+}=\sum_{l,m=0}^{\infty}W(f)^{*}\Xi_{l,m}(\kappa_{-+}(l,m;f)).\]
 Since $W(f)^{*}\Xi_{--}W(f)\in\mathcal{L}(\mathcal{E}_{+},\mathcal{E}_{+}^{*})$,
there exists a unique kernel distribution $\kappa_{--}(l,m;f)$ satisfying\[
\Xi_{--}=\left(W(f)^{*}\right)^{2}\Xi_{--}W(f)^{2}=\sum_{l,m=0}^{\infty}W(f)^{*}\Xi_{l,m}(\kappa_{--}(l,m;f))W(f).\]
For $\Xi_{++}\in\mathcal{L}(\mathcal{E}_{+},\mathcal{E}_{+}^{*})$,
we also have a unique kernel distribution $\kappa_{++}(l,m)$ satisfying\[
\Xi_{++}=\sum_{l,m=0}^{\infty}\Xi_{l,m}(\kappa_{++}(l,m)).\]

Therefore we obtain \[
\Xi=\sum_{l,m=0}^{\infty}\Xi_{l,m},\]
where\begin{align*}
\Xi_{l,m} & =\Xi_{l,m}(\kappa_{++}(l,m))+\Xi_{l,m}(\kappa_{+-}(l,m;f))W(f)\\
 & \quad+W(f)^{*}\Xi_{l,m}(\kappa_{-+}(l,m;f))+W(f)^{*}\Xi_{l,m}(\kappa_{--}(l,m))W(f).\end{align*}

\end{proof}

\end{document}